\documentclass[number,sort&compress,5p]{elsarticle}
\pdfoutput=1

\usepackage{multirow}
\usepackage{graphicx}% Include figure files
\usepackage[version=3]{mhchem}
\usepackage{upgreek}
\usepackage{url}
\usepackage{verbatim}
\usepackage{fancyvrb}
\VerbatimFootnotes
\usepackage{epstopdf}

\title{Window Function for Chirped Pulse Spectroscopy with Enhanced Signal-to-noise Ratio and Lineshape Correction}

\author{Luyao Zou}
\author{Roman A.\ Motiyenko\corref{cor1}}
\ead{roman.motiyenko@univ-lille.fr}
\cortext[cor1]{Corresponding author}
\address{Univ. Lille, CNRS, UMR 8523 - PhLAM - Physique des Lasers Atomes et Mol\'{e}cules, 59000 Lille, France
}%

\begin{document}

\begin{abstract}

In chirped pulse experiments, magnitude Fourier transform is used to generate frequency domain spectra.
The application of window function as a tool for lineshape correction and signal-to-noise ratio (SnR) enhancement 
is rarely discussed in chirped spectroscopy, with the only exception of using Kaiser-Bessel window and trivial rectangular window. 
We present a specific window function, called ``Voigt-1D'' window, designed for chirped pulse spectroscopy.
The window function corrects the magnitude Fourier-transform spectra to Voigt lineshape, 
and offers wide tunability to control the SnR and lineshape of the final spectral lines. 
We derived the mathematical properties of the window function, 
and evaluated the performance of the window function in comparison to the Kaiser-Bessel window on experimental and simulated data sets. 
Our result shows that, compared with un-windowed spectra, the Voigt-1D window is able to produce 100~\% SnR enhancement on average.
\end{abstract}


\maketitle

\section{Introduction}

Fourier-transform has been widely used in spectroscopy, such as nuclear magnetic resonance (NMR)\cite{Ernst1966RSI}, 
Fourier-transform microwave spectroscopy\cite{Balle1981RevSciInstrum}, and the Fourier transform chirped pulse spectroscopy\cite{Pate2008CPMW}.
Standard signal processing offers a wide range of tools, 
such as windowing, filtering, re-sampling, and wavelet transforms, to extract the spectral information from a noisy signal.
Signal to noise ratio (SnR) and spectral resolution are the two main characteristics that a spectroscopist would be concerned about. 
For an exponential free-induction decay (FID) signal of lifetime $\tau$ and data acquisition (DAQ) length $t_N$, 
the SnR is proportional to $(1-e^{-t_N/\tau})\tau/\sqrt{t_N}$, and the maximum SnR is achieved at $t_\text{max}\approx 1.26\tau$.\cite{Matson1977JMR} 
Window functions are used to go beyond this intrinsic SnR -- DAQ length relationship.
Applying a window function to the FID can be viewed as applying different weighting factors on different regions of the FID, 
thus emphasizing the regions with stronger signal for higher SnR. \cite{WinFuncNMR}
There is often compromise, however, between achieving high SnR and high spectral resolution of a line, 
because applying window function in the time domain is equivalent to convolving the Fourier transform of the window function with the spectral line 
in the frequency domain, which broadens the spectral line. 
Combined with window functions, more sophisticated methods based on non-uniform\cite{Palmer2015JPCB} sampling have been applied, 
mostly in multi-dimensional NMR, to further increase the SnR without losing spectral resolution.\cite{Simon2019JBioNMR, Kaur2020JPCA}

The invention of chirped pulse Fourier-transform (CP-FT) microwave spectroscopy \cite{Pate2008CPMW} allowed extremely rapid and broadband up to several tens of GHz acquisition of a rotational spectrum. Due to this double advantage, CP-FT spectroscopy has become actively used for structural determinations in large and complex molecular systems (large molecules and clusters) \cite{Perez2013CPL, Uriarte2019}, for reaction dynamics and kinetics studies \cite{dian2008measuring, prozument2014, hays2020}, and for rapid chemical composition analysis \cite{Gerecht2011OptExp, crabtree2016microwave}. See also a review paper by Park \& Field \cite{park2016perspective}. The former requires measurements of heavy atoms isotopic species often in natural abundance that are represented by weak satellite spectrum in the vicinity of strong parent isotopic species lines. Although high SnR is a major goal for detecting weak lines, the spectral resolution at baseline is often considered as the most important parameter in the molecular structure determination \cite{Perez2013CPL}. The studies of chemical reaction dynamics and kinetics requires the determination of accurate spectral lineshape parameters (may be also obtained from the analysis of the time domain FID) \cite{hays2020}, and high SnR to detect weak lines of rare or highly excited reaction products \cite{prozument2014}. Finally, the chemical composition analysis is mostly focused on obtaining highest possible SnR to detect ppm and ppb concentrations \cite{Gerecht2011OptExp}. In addition to the above mentioned examples, one of the primary applications of rotational and thus of CP-FT spectroscopy is a spectral characterization of new molecular species aiming for the determination of its Hamiltonian parameters. Such study would require high line frequency measurement accuracy which depends both on SnR and spectral resolution usually limited by spectral line full-width at half-maximum (FWHM). Whereas very high spectral resolution may be achieved in the microwaves using molecular beam experiments due to effectively cooled molecules, the Doppler broadening becomes important in millimeter and submillimeter-wave measurements, and in general in room-temperature spectra.

In contrast to NMR, few studies are focused on the SnR enhancement for chirped pulse spectroscopy. 
Although similar to NMR in the fundamental mechanism, FID signals in chirped pulse spectroscopy are often more complicated than those in NMR. 
Firstly, the chirped pulse excitation, which is a rapid linear sweep of radiation frequency of a large bandwidth,
induces multiple molecular transitions, each of which occurs at a slightly different time during the chirp. 
This results in a mixture of FID signals with different initial phases that prohibit one from performing
a uniform phase correction and then extract the real part of the Fourier transform, as routinely used by the NMR community.
Instead, chirped pulse spectroscopists usually use the magnitude Fourier transform of the FID directly.
which combines the real (absorption) and imaginary (dispersion) part of the Fourier transform.
Although magnitude Fourier transform does not shift the line position, 
it increases the linewidth of the Fourier transform spectra and creates wide wings on the line profile, because of the extra contribution from the dispersion.
Mathematically, it means the line profile changes from a ``real Voigt'' (real part of the Faddeeva function) profile 
to a ``complex Voigt'' (the magnitude of the Faddeeva function) profile.
Line asymmetry due to initial phase is also amplified by the complex Voigt profile because of its large wings.
Secondly, compared to NMR, chirped pulse spectroscopy, especially in the  (sub)millimeter regime, has a much shorter FID that is a hybrid of exponential and Gaussian decays, 
due to the increasingly significant Doppler broadening at higher transition frequencies.
The much shorter FID means that the delay between the pulse excitation and DAQ start time becomes non-negligible. 

As a simple yet effective tool, 
appropriate window functions are expected to correct the initial phase of the chirp FID and therefore generate symmetric spectral lines. 
In chirped pulse spectroscopy, however, only a few examples clearly discussed the effect of window functions\cite{Pate2008CPMW, Steber2012JMS}.
Other studies specified the usage of window function\cite{Neill2013OptExp, Suits2014JCP2, Fatima2020PCCP} without further discussion. 
In these studies, the standard, symmetric Kaiser-Bessel window (short for Kaiser window hereafter) is the common choice, 
because of its known capability of sidelobe suppression.
The SnR, however, is sacrificed. 
Square window is also mentioned in several studies \cite{Pate2008CPMW, Steber2012JMS, hays2020},
but it is trivial since adjusting a square window is equivalent to adjusting the on-site DAQ time settings.
The Kaiser window offers limited flexibility, with only one parameter $\pi\alpha$ that adjusts the width of the window function. 
When $\pi\alpha \rightarrow 0$, it becomes a square window, and when $\pi\alpha \rightarrow +\infty$, it approximates a Gaussian window.
In the chirped pulse literature that mentioned the use of Kaiser window, 
$\pi\alpha$ is set to 8 \cite{Pate2008CPMW, Steber2012JMS, Fatima2020PCCP},
a value that produces nearly Gaussian lineshape in the frequency domain. 

In order to obtain wider tunability on the spectral SnR and lineshape, 
we propose an asymmetric window function specifically designed for treating chirped pulse data, 
denoted as the ``Voigt-1D'' window.
The window function is inspired by the concept of ``matched window''\cite{Turin1960IRE}.
In this paper, optimal window parameters are chosen to achieve the maximum SnR at the highest possible spectral resolution. 
The proposed window function is able to improve SnR of the Fourier transformed spectral line, and also correct the lineshape profile of the magnitude spectrum from complex Voigt to real Voigt. 
For simplicity, when ``Voigt profile'' is used alone without specifying ``real'' or ``complex'' hereafter, it refers to the ``real Voigt'' profile. 
The window function includes two tuning parameters that can offer much larger flexibility than standard symmetric window functions. 
In this article, we present the window function and its mathematical properties, 
derive the SnR expression with respect to its tuning parameters, 
and then show its performance based on both simulated data and real experimental data. 
We also suggest a general guideline for choosing the window function parameters,
so that one can obtain close-to-optimal SnR and spectral resolution based on the properties of the chirped pulse dataset. 
% In practice, one can choose to manually adjust the window parameters for specific goals, 
% and can also program using the guideline for automated data-processing tasks.  

\section{Mathematics of the Voigt-1D window}

\subsection{The window function and its property}

The Voigt-1D window is named after ``Voigt'' and ``1D''.
``Voigt'' means that it is a product of a Gaussian decay and a Lorentzian decay, which produces a Voigt profile after Fourier transform. 
``1D'' means it is related to the 1\textsuperscript{st} derivative of the Voigt profile.

\begin{figure}[htp!]
\centering 
\includegraphics{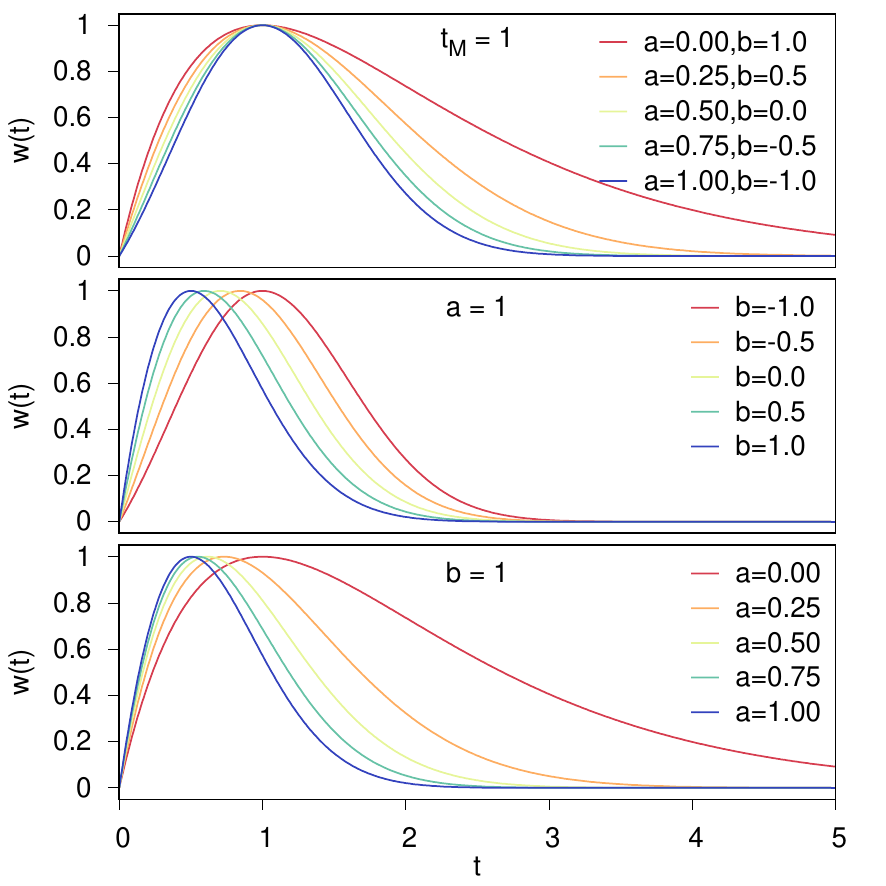}
\caption{The shape of a Voigt-1D function. (top) The maximum value is fixed at $t_\text{M}=1$; (center) $a$ is fixed at 1; (bottom) $b$ is fixed at 1. \label{fig:winf-shape}}
\end{figure}

The Voigt-1D window combines the concept of matched window, and the ability for the FID initial phase correction that removes spectral leakage and corrects spectral lineshape.
If only white Gaussian noise is present in the time domain signal, a ``matched window'' takes the same function form of the FID envelope.
We can write the envelope of an FID from a single molecular line as 
\begin{equation}
f(t; a_0, b_0) = \exp\left(-a_0 t^2 - b_0 t\right) \quad (t\geq 0) \label{eq:FID}
\end{equation}
where $t$ represents the elapsed time from the start point of DAQ,
and $a_0$ and $b_0$ are coefficients that correspond to the Gaussian and exponential components of the decay, respectively.
Following Eq.~\ref{eq:FID}, the Voigt-1D window is defined as 
\begin{equation}
w(t; a, b) = \frac{t}{M(a,b)}\exp\big(-at^2-bt\big) \quad (t\geq 0) \label{eq:winf}
\end{equation}
where $a$ and $b$ are the two tuning parameters, and $M(a,b)$ is the normalization factor that sets $\text{max}\{w(t)\} = 1$. 
To ensure $w(t) \rightarrow 0 (t \rightarrow 0)$, $a$ must be non-negative. 
Figure~\ref{fig:winf-shape} shows the shape of $w(t)$ with a selective set of parameters. 

The Voigt-1D window function is non-negative. 
As $t$ increases, the function first monotonically increases from 0, followed by an asymptotic approach back to 0. 
For a given pair of $a$ and $b$, the maximum of $w(t)$ is reached at 
\begin{equation}
t_\text{M}=
 \begin{cases}
 \cfrac{\sqrt{b^2 + 8a} - b}{4a} ,\quad (a > 0) \\
 \cfrac{1}{b}, \quad (a=0)
 \end{cases}
\label{eq:max-winf-t}
\end{equation}
Consequently, 
\begin{equation}
\begin{gathered}
M(a, b) = \\
  \begin{cases}
  \cfrac{4a}{\sqrt{b^2+8a}-b}\exp\Big(\cfrac{1}{2} + \cfrac{b\sqrt{b^2+8a}-b^2}{8a}\Big), \quad (a>0)\\
  \cfrac{1}{be}, \quad (a=0)
  \end{cases}
\label{eq:M-ab}    
\end{gathered}
\end{equation}

The Voigt-1D window has the following properties:
\begin{itemize}

\item The window function is the FID envelope $f(t)$ multiplied by $t$. 
The Fourier transform of $w(t)$ is $(i/2\pi) \hat{f}'(x)$, where $\hat{f}(x)$ is the Fourier transform of the $f(t)$ , 
and $\hat{f}'(x)$ is its first derivative. Using this relationship, we can show that the lineshape of the magnitude Fourier transform spectrum of the FID, 
after being multiplied with the window function, is exactly Voigt (see supplementary material, Section~1). 
This property ensures the ability of lineshape correction of the window function,
except for extremely negative $b$ ($b < -2\sqrt{a}$), where sidelobes start to appear in the lineshape profile. 

\item The product of the window function $w(t;a,b)$ and an FID envelope $f(t;a_0, b_0)$ is still in the form of the window function $w(t; a', b')$,
where parameters are updated as $a'=a+a_0$, and $b'=b+b_0$. That is, 
\begin{equation*}
\begin{gathered}
w(t;a, b)f(t; a_0, b_0) = \\
\frac{M(a+a_0, b+b_0)}{M(a, b)}w(t; a+a_0, b+b_0)
\end{gathered}
\end{equation*}

\item The ratio between $a$ and $b$ determines the shape of the window function. For a given $t_\text{M}$, $a$ increases as $b$ decreases, 
and larger $a$ generates a narrower window shape that approaches a Gaussian window. 
For a given $a$ or a given $b$, the increase of the other parameter result in a smaller $t_\text{M}$ and also narrower window. 
This property shows the wide tunability of the window function.

\item The window function is independent of the FID length. In contrast to standard symmetric window functions, 
whose maximum is always at the center of the FID, the Voigt-1D window simply extends its values when the FID length is extended. 
This property ensures the result of windowing is not affected by experimental adjustments on the DAQ length, 
making it ideal for automatic processing of segmented chirped pulse data that cover multiple frequency bands. 

\end{itemize}

\subsection{Metrics for performance measurement}

As widely accepted metrics in spectroscopy, SnR and FWHM characterize the intensity and the resolution of spectral lines, respectively.
High SnR and narrow FWHM are often anti-correlated, which cannot be achieved simultaneously. 
We choose SnR/FWHM, in addition to SnR and FWHM, to measure the performance of the Voigt-1D window, 
so that maximizing SnR/FWHM reflects the goal of reaching highest possible SnR with highest possible spectral resolution.

Another implicit performance metric is the sampling density of the frequency grid, $\Delta\nu$, 
which is the reciprocal of the length of the time domain signal.
In discrete Fourier transform, $\Delta\nu$ can be improved by zero-padding, i.e.,
appending certain length of $0$ arrays to the end of the original signal.
The side-effect of zero-padding is to cause periodic ripples on the baseline due to spectral leakage.

\subsection{Evaluation of the performance metrics: SnR \label{section:eval-snr}}

The SnR is evaluated by the ratio of the peak intensity and the noise level of the spectrum. 
These quantities can be calculated directly from mathematical formulae for any FID signal. 

The peak intensity of an FID signal $f(t)$ from $t=0$ to $T$ is $\int_0^T f(t)\mathrm{d}t$.
If the signal is multiplied by a window function $w(t)$, the peak intensity becomes 
$\int_0^T f(t)w(t)\mathrm{d}t$.

Assume the noise in the time domain is a pure white Gaussian noise with standard deviation of $\sigma$. 
The standard deviation of the magnitude Fourier transform of the noise, from $t=0$ to $T$, is
$\sqrt{T} \sigma$. 
If the time domain white Gaussian noise is multiplied by a window function $w(t)$, 
the frequency domain noise has a standard deviation of $\sigma\sqrt{\int_0^{T} w^2(t)\mathrm{d}t }$ 
(see supplementary material, Section~2). 
For simplicity, we may assume $\sigma=1$. 
We may see that the noise of un-windowed FID grows to infinity as $T\rightarrow +\infty$, 
and therefore the upper bound $T$ needs to be specified for calculating the SnR.
On the other hand, for a window function of which $\int_0^{+\infty} w^2(t)\mathrm{d}t $ is finite, 
the noise is also finite even when  $T\rightarrow +\infty$.

Using the equations above, we can write the SnR of a windowed spectral line as 
\begin{equation}
\text{SnR}(T) = \frac{\int_0^T f(t)w(t)\mathrm{d}t}{\sqrt{\int_0^T w^2(t)\mathrm{d}t}} \label{eq:snr-general}
\end{equation}

From Equation~\ref{eq:snr-general}, the SnR of a un-windowed spectral line is
\begin{equation}
\begin{aligned}
& \text{SnR}(T) = \frac{\int_0^T f(t; a_0, b_0)\mathrm{d}t}{\sqrt{T}} = \\ 
& 
\begin{cases}
  \cfrac{1 - e^{-b_0T}}{b_0\sqrt{T}}, \quad (a_0=0) \label{eq:snr-un-windowed}\\
  \cfrac{\sqrt{\pi}}{\sqrt{4a_0T}}e^{b_0^2/4a_0} \bigg[\text{erf}\Big(\cfrac{2a_0T+b_0}{\sqrt{4a_0}}\Big)  - \text{erf}\Big(\cfrac{b_0}{\sqrt{4a_0}}\Big)\bigg], (a_0>0)
\end{cases}
\end{aligned}
\end{equation}
where $\text{erf}(x)=\cfrac{2}{\sqrt{\pi}}\int_0^x e^{-t^2}\mathrm{d}t$ is the error function.
The maximum SnR is reached by finding the root of $\text{SnR}'(T)=0$. In the $a_0=0$ case, 
the numerical result is $t_\text{max}\approx 1.26/b_0$, which was stated by Matson \cite{Matson1977JMR}.
The maximum SnR at $t_\text{max}$ is approximately 0.715 (the $a_0=0$ case). 

For the Voigt-1D window defined in Eq.~\ref{eq:winf}, 
we consider the $T\rightarrow +\infty$ case because $\int_0^{+\infty} w^2(t)\mathrm{d}t$ is finite.
We define the following two integrals:
\begin{align}
P(a, b) &= \int_0^{+\infty} t\exp(-at^2-bt)\mathrm{d}t  \label{eq:intg-Pab-inf}\\ 
Q(a, b) &= \int_0^{+\infty} t^2\exp(-2at^2-2bt)\mathrm{d}t \label{eq:intg-Qab-inf}
\end{align}
In the general case where $a > 0$, 
\begin{align}
P(a,b) &= \frac{1}{2a} - \frac{b\sqrt{\pi}}{4\sqrt{a^3}}\text{erfcx}\Big(\frac{b}{2\sqrt{a}}\Big) \label{eq:intg-Pab} \\
Q(a,b) &= \frac{\sqrt{2\pi}(a+b^2)}{16\sqrt{a^5}}\text{erfcx}\Big(\frac{b}{\sqrt{2a}}\Big) 
- \frac{b}{8a^2} \label{eq:intg-Qab}
\end{align}
where $\text{erfcx}(x) = e^{x^2}(1-\text{erf}(x))$ is the scaled complementary error function.
In the Lorentzian-limited case where $a=0$,
\begin{align}
P(0, b) &= \int_0^{+\infty} te^{-bt}\mathrm{d}t = \frac{1}{b^2} \label{eq:intg-Pab-limit} \\
Q(0, b) &= \int_0^{+\infty} t^2e^{-2bt}\mathrm{d}t = \frac{1}{4b^3} \label{eq:intg-Qab-limit}
\end{align}

With these two integrals, we can write 
the SnR of a Voigt-1D windowed spectral line, 
with FID decay parameters $(a_0, b_0)$ and window parameters $(a, b)$, as
\begin{equation}
s(a_0, b_0, a, b) = \frac{P(a_0+a, b_0+b)}{\sqrt{Q(a, b)}} \label{eq:snr-theory}
\end{equation}

To find the maximum point of $s(a_0, b_0, a, b)$ with respect to $a$ and $b$, 
numerical method is necessary because its partial derivatives are transcendental functions that can only be solved numerically. 
In our computer code, we used standard least-square method to find the minimum of the negation of $s(a_0, b_0, a, b)$ 

\subsection{Evaluation of the performance metrics: FWHM}

The full widths of the spectral lines can only be calculated numerically. 
To find the full widths at level $y$, we simulated the spectral profile $f(x)$, 
find the numerical points that are closest to $f(x)=y$, 
and then measure the distance between these points. 
For un-windowed and Voigt-1D windowed spectra, we can write out the exact equations of $f(x)$, 
and therefore can simulate the spectral profile with a fine $x$ grid numerically. 
For Kaiser windowed spectra, we do not have the exact lineshape function, 
and therefore the spectral profile is generated by digital Fourier transform, and smoothed by univariate spline interpolation.

\subsection{Determination of FID initial parameters $a_0$ and $b_0$}

\begin{figure}[htp!]
\includegraphics{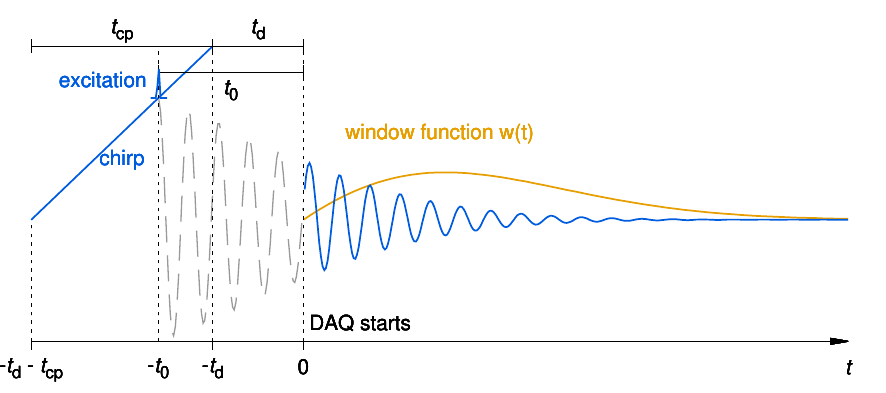}
\caption{Illustration of the reference frame of time in a chirped pulse experiment. \label{fig:time-frame}}
\end{figure}

The value of the performance metrics are not only the function of the window function,
but also the shape of the original FID that is controlled by the initial parameters $a_0$ and $b_0$. 
In the derivations above, we set the start of DAQ as $t=0$. 
This frame of reference simplifies the mathematical expressions of the window function and performance metrics. 
In real chirped pulse experiments, however, the chirp excitation has a noticeable duration $t_\text{cp}$.
In addition, a certain dead time $t_\text{d}$ is often placed between the end of excitation and the start of DAQ, 
so that electronics can be correctly switch on or off, and transient noises can be blocked. 
Figure~\ref{fig:time-frame} illustrates the relations between these times.
In this frame of reference, the actual start of the FID of a molecular line is $-t_0$ ($t_0>0$) in Figure~\ref{fig:time-frame},
which is between $-t_\text{d}$ and $-t_\text{d}-t_\text{cp}$ and can be calculated if the excitation frequency is known (see Supplementary Material, Section 3.6). 

The presence of $t_0$ affects the shape of the FID envelope. 
Let $b_\lambda$ to be the real Lorentzian component of the FID, we may write the FID envelope as 
\begin{equation}
\begin{aligned}
f(t) &= \exp(-a_0(t+t_0)^2 - b_\lambda(t+t_0)) \\
&= \exp(-a_0 t^2 - (b_\lambda + 2a_0t_0) t) \cdot \exp(-a_0t_0^2-b_\lambda t_0)
\end{aligned}
\end{equation}
Compared with Equation~\ref{eq:FID}, we have $b_0 = b_\lambda + 2a_0t_0$, i.e., 
the exponential decay term depends not only on the real broadening effect due to physics, 
but also on the Gaussian decay term and delay time $t_0$ due to mathematical relations. 
The extra intensity scalar $\exp(-a_0t_0^2-b_\lambda t_0)$ represents the loss of line intensity due to $t_0$, 
which is discussed by Gerecht et al.\cite{Gerecht2011OptExp}.
The effect of $t_0$ is independent of the choice of frame of reference. 

When the FID is Lorentzian dominated, i.e., $b_\lambda \geq a_0$, the effect of $t_0$ is not obvious. 
However, when the FID is Doppler dominated, the effect can be significant. 
Doppler dominated FID occurs both in room-temperature (sub)millimeter chirped pulse spectroscopy, 
where the Doppler broadening of the molecular line is not negligible, 
and in jet cooled spectroscopy, both microwave and (sub)millimeter, where the residual Doppler components from the jet is the main dephasing mechanism.

The optimization of Voigt-1D window function parameters depends on the FID initial parameters $a_0$ and $b_0$, 
and therefore implicitly depends on $t_0$. 
For a narrowband chirp, or single line excitation, $t_0$ can be precisely calculated and therefore this effect 
can be precisely accounted for in Equation~\ref{eq:snr-theory}. 
For a broadband chirp, each line has its unique $t_0$ that spans the whole chirp excitation, 
and this effect is especially significant when $a_0t_0\geq b_\lambda$. 
A window function, however, has an explicit shape that cannot be the optimal choice for all the $t_0$ possibilities. 
It is inevitable that compromise has to be made, and in this paper, we choose to use the center point of the chirp excitation, 
$t_\text{d}+t_\text{cp}/2$, as the ``averaged'' $t_0$ to determine the window function parameters.

\subsection{Unit convention}

In deriving the mathematical expressions, we did not specify the units of $t$, $a$, and $b$. 
$a$ has the dimension of time$^{-2}$, and $b$ has the dimension of time$^{-1}$.
Our mathematical derivations can be rescaled to any unit system without altering their properties, 
as long as $at^2$ and $bt$ are kept dimensionless, and the ratio $b/2\sqrt{a}$ remains unchanged. 
In our data treatment, we use $\upmu$s for $t$, and therefore $a$ is in MHz$^2$ and $b$ is in MHz. 

\section{Results and Discussion}

\subsection{Choice of window parameters based on theoretical SnR \label{section:choice-of-param}}

\begin{figure*}[htp!]
\centering
\includegraphics[width=\textwidth]{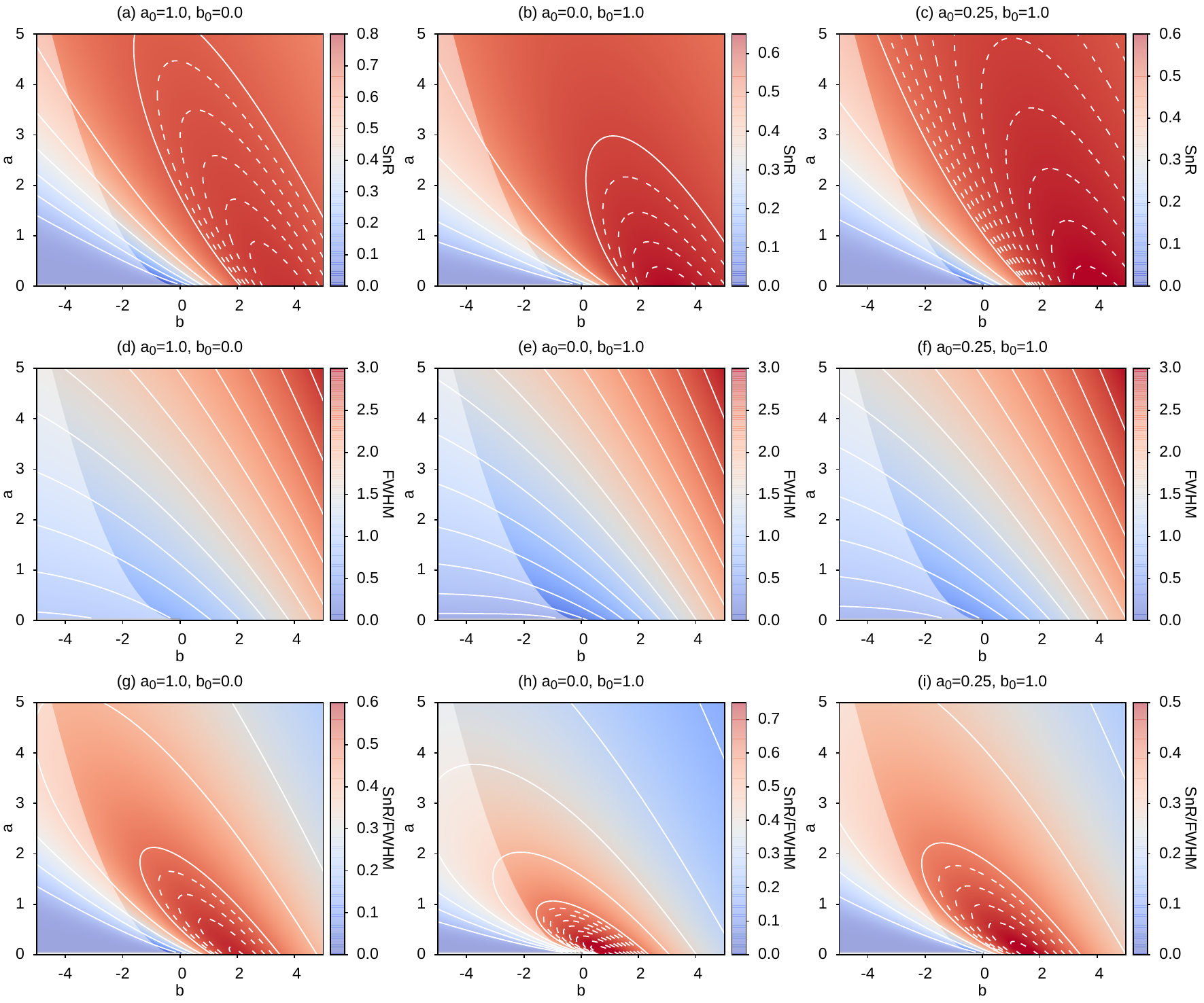} % remove width=... for 2-column version !!!!!! 
\caption{Theoretical SnR (top row), FWHM $\gamma_V$ (center row), and SnR/FWHM (bottom row) at the $T\rightarrow +\infty$ limit.
(a,d,g) $a_0=1.0, b_0=0$, (b,e,h) $a_0=0, b_0=1.0$, (c,f,i) $a_0=0.25, b_0=1.0, b = 2\sqrt{a}$.
In the SnR and SnR/FWHM plots, the solid contour lines are spaced by 0.1, and the dashed contour lines are spaced by 0.01
For the FWHM plots, the solid contour lines are spaced by 0.2.
The shaded areas in all plots mark the unfavored negative $b$ values where $b < -2\sqrt{a}$. 
\label{fig:snr-theory}}
\end{figure*}

Figure~\ref{fig:snr-theory} visualizes Eq.~\ref{eq:snr-theory} and its corresponding FWHM as a function of $a$ and $b$ for given $(a_0, b_0)$ pairs.
We calculated three sets: $(a_0, b_0)=(1, 0)$ for pure Gaussian decay, $(a_0, b_0)=(0, 1)$ for pure exponential decay, and an intermediate case $(a_0, b_0)=(0.25, 1)$, so that $b = 2\sqrt{a}$. 
Interestingly,  the maxima of both SnR and SnR/FWHM are reached when $a=0$, i.e., 
the Voigt-1D function is in fact a pure first derivative of the Lorentzian. 
It can be justified by two reasons.
First, when $a=0$, the area under the window function is the largest (see Figure~\ref{fig:winf-shape}), 
and therefore the window function collects the most signal. 
Second, the FWHM is smaller when $a=0$ because the window function maximizes the Lorentzian component in the line profile.
The $b$ value for maximum SnR/FWHM is approximately half of the $b$ value for maximum SnR.

Under the guide of Eq.~\ref{eq:snr-theory} and the numerical methods used to generate Figure~\ref{fig:snr-theory},
one can find an optimal $(a, b)$ parameter set of the Voigt-1D window for a particular FID signal, 
according to the metric to be optimized. 
To find this parameter set, the initial parameters $(a_0, b_0)$, in other words, 
the Gaussian ($\gamma_G$) and Lorentzian ($\gamma_L$) FWHM of the original FID signal,
need to be determined. 
If fitting the envelope of the FID is convenient, one can retrieve $(a_0, b_0)$ from the time domain fit. 
If not, fitting a complex Voigt profile to the un-windowed frequency domain spectrum may
also lead to a reasonable estimation of $a$ and $b$. 
Especially, \textit{a priori} one may fix $a$ to $\pi^2\gamma_G^2/(4\ln2)$, 
assuming that $a$ is completely caused by Doppler broadening of the molecule. 
Although no analytical expression of obtaining the optimal $(a, b)$ parameters can be specified, 
this procedure to find optimal parameter sets can be easily programmed or tabulated to allow automatic processes.

Figure~\ref{fig:snr-theory} also demonstrates the wide tunability of the Voigt-1D window function.
Around the maxima, there is a wide range of $a$ and $b$ that can yield SnR and SnR/FWHM values more than 90~\% of the theoretical maxima. 
At a given SnR, a series of $(a, b)$ pairs can be chosen to produce different FWHM.
In turn, $(a, b)$ pairs on the same contour line in the center row of Figure~\ref{fig:snr-theory} 
produce identical FWHM with different weights of the Gaussian and Lorentzian components. 
Such tunability is helpful when we apply the window function to broadband spectra, 
where the Doppler component intrinsically varies across the whole spectra.
We can show that using the ``averaged'' window parameters derived from the ``averaged'' $t_0$ can produce
SnR and SnR/FWHM near the optimal values across the broadband chirps (See Supplementary Material, Section 4.1).

The flexibility of the Voigt-1D window allows the user to tune the window away from its ``optimal'' performance. 
For example, if high resolution is the priority, both $a$ and $b$ need to be as small as possible, 
because the line width metrics always monotonously increases as $b$ increases for any given $a$, and as $a$ increases for any given $b$. 
$a=a_0$ and $b=-2\sqrt{a_0}$ can be a good initial parameter set, and one may further adjust $a$ and $b$ following the constraint $b=-2\sqrt{a}$ 
for maintaining good lineshape, until a satisfactory resolution is achieved. 
In this scenario, negative $b$ does not infer anything about an exponentially increasing FID, which is impossible, 
but servers only as a mathematical treatment to obtain narrow linewidths, at the cost of losing SnR.

\subsection{SnR enhancement on single-peak spectra}

The performance of the Voigt-1D window was first tested on a set of OCS lines, consisting of 51 entries. 
These lines were experimentally measured using a millimeter wave chirped pulse spectrometer\cite{Zou2020RSI} (See Supplementary Material, Section 3.1 for experimental details).
In this data set, each line is the only line within its chirp bandwidth, so that the {$(a_0, b_0)$ parameters of the FID envelope can be unambiguously modeled. 
We chose the lines from the \ce{OC^{34}S} isotopologue and the $v_2=1$ vibrational excited state of the parent isotopic species, 
in order to avoid saturation effects on the ground state lines of the latter. 
The line frequencies range from the 60~GHz to 300~GHz, 
where the lineshape shifts from Lorentzian-dominated to nearly equal weight of Lorentzian and Gaussian components.

For each entry, we treated the data with 2 sets of Voigt-1D window parameters, set (1) optimized for maximizing SnR, 
and set (2) optimized for maximizing SnR/FWHM. 
Before applying the Voigt-1D window, we retrieved the initial $a_0$ and $b_0$ values of the line by fitting the time domain FID signal 
(see supplementary material, Section~3.4).
In the fit, $a_0$ is fixed to $\pi^2\gamma_G^2 /(4\ln2)$, where $\gamma_G$ is the Doppler FWHM of the OCS line, and $b_0$ is fitted from the FID. 
Afterwards, the two parameter sets were numerically solved by fixing $a=0$, as demonstrated in Figure~\ref{fig:snr-theory}.
The results were compared with 3 sets of Kaiser window $\pi\alpha$ values, 0, 4, and 8. 
$\pi\alpha=0$ is equivalent to un-windowed spectrum, $\pi\alpha=8$ is heavily windowed for best baseline resolution, 
and then $\pi\alpha=4$ is the medium state between un-windowed and heavily windowed spectra. 
Before applying the Kaiser windows, the FID signal was truncated to the length that maximizes theoretical SnR of the un-windowed spectrum (Equation~\ref{eq:snr-un-windowed}).

Figure~\ref{fig:ocs-summary} shows the SnR (top panel), FWHM (center panel), and SnR/FWHM (bottom panel) of all 51 entries.
In all three plots, the results of the un-windowed spectra, i.e., Kaiser $\pi\alpha=0$, are shown as gray bars in the background,
providing the reference point for comparison.
The enhancement of the four windows are plotted as line series. 
100~\% means that the value to be compared (SnR, FWHM, or SnR/FWHM) of the windowed line is identical to the un-windowed line. 
Note that the 51 entries are independent samples, and therefore the line series are only for visual distinction and does not indicate any correlation between entries.

The performance of four tested windows are summarized as the following:

(1) The SnR. For almost all entries, the Voigt-1D window produces spectral lines of higher SnR than the un-windowed lines, which are already of the highest SnR in theory.
This is because the noise in the real FID signals is not pure white noise, and may vary from shot to shot. 
The Voigt-1D window, however, is insensitive to truncation and therefore can be applied to a long FID record and collect 
as much information as possible.
Two parameter sets of the Voigt-1D window result in similar SnR enhancements, 
which corresponds to the large tunable parameter space shown in Figure~\ref{fig:snr-theory}.
Kaiser windows also improve the SnR in most cases, but the enhancement is less than that from Voigt-1D windows. 
The higher SnR enhancement by Voigt-1D window can be explained because the Voigt-1D window does not truncate data. 
The window shape weights the data points such that the noise is suppressed, but it still uses all the information of the full data series. 
The Kaiser window, on the other hand, has to discard the data after a certain length in exchange for higher SnR, 
because the Kaiser window has a rigid symmetric window shape.

(2) The FWHM. For all entries, the un-windowed spectra present the smallest FWHM. 
It is not surprising because applying window function in the time domain is equivalent to convolving the window function profile with the spectral line profile. 
Nevertheless, the Voigt-1D set (2) only broadens the line by approximately 25~\%. 
These values demonstrate that the Voigt-1D set (2) window does not severely broaden the line in exchange for SnR enhancement.
The FWHM produced by Voigt-1D set (1) is similar to that produced by Kaiser window with $\pi\alpha=4$.
The FWHM is almost doubled. 
Finally, Kaiser window with $\pi\alpha=8$ produces 2.5 times broader FWHM. 

(3) The SnR/FWHM. Following the discussion of SnR and FWHM, 
it is expected that Voigt-1D set (2) produces the optimal SnR/FWHM among the 4 tested windows.
This window proves that a slight line broadening in exchange for a significant SnR improvement is feasible. 
Voigt-1D set (1) is in second place, and has higher SnR/FWHM than Kaiser $\pi\alpha=4$,
because they produce similar FWHMs but Voigt-1D set (1) produces higher SnR. 
The Kaiser $\pi\alpha=8$ window has disadvantage in this SnR/FWHM metric, because this window significantly broadens the spectral lines. 
Nevertheless, Kaiser $\pi\alpha=8$ will have better result if the focus is on baseline resolution,
which is important in some applications, such as the identification of isotopologue species in their natural abundances\cite{Perez2013CPL}.

In addition, applying the Voigt-1D window can help improving frequency sampling together with the SnR enhancement, 
because of its insensitivity to zero-padding. 
In the tests discussed above, the high SnR spectra were obtained by FID truncation, and no zero-padding was applied. 
There are only 3--5 points to describe a line in the frequency domain spectrum. 
To improve frequency sampling, extending the FID or zero-padding is necessary. 
Shown in Equations~\ref{eq:intg-Pab-inf}--\ref{eq:snr-theory}, Voigt-1D window does not lose SnR in zero-padding. 
Also, since the window function starts at 0, it suppresses the baseline ripples due to spectral leakage. 
These ripples, although are not real noise in nature, contaminates the spectral baseline and produces effective larger noise in our SnR calculation. 
Examples of such ripples can be found in Figure~S2.
Therefore, if zero-padding is used, the SnR of the un-windowed spectra shown in Figure~\ref{fig:ocs-summary} will be lower, 
and the SnR enhancement by Voigt-1D window calculated in this way will be higher than what has been presented. 
The Kaiser window also suppresses these ripples, but only with sufficient large $\pi\alpha$ values (e.g., $>5.6$), 
which broadens the FWHM. 

\begin{figure}[htp!]
\centering 
\includegraphics{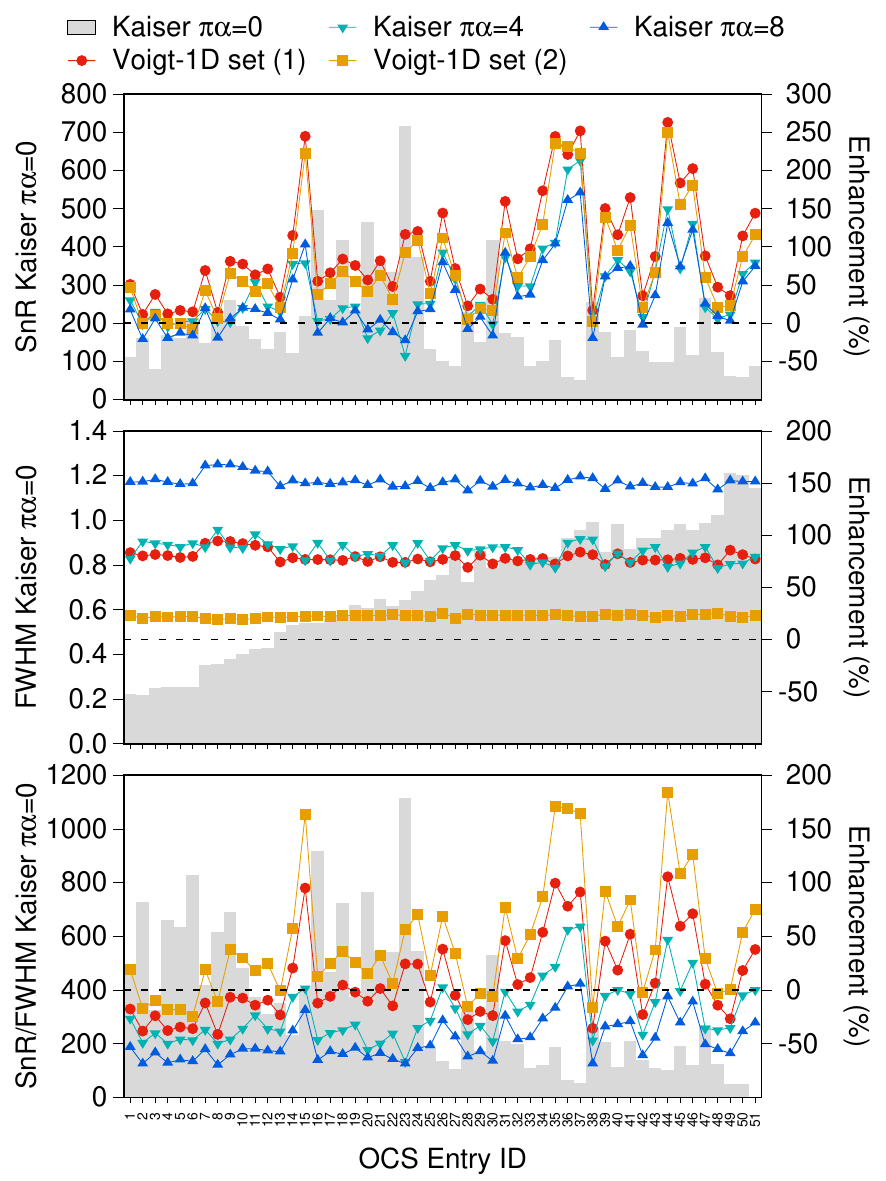}
\caption{
Summary of SnR (top), FWHM (center), and SnR/FWHM (bottom) of 51 OCS lines, treated with Voigt-1D windows and Kaiser windows. 
For Voigt-1D windows, parameter set (1) was set to maximizing SnR, and set (2) was set to maximize SnR/FWHM. 
For Kaiser windows ($\pi\alpha=0, 4, 8$), the FID signal is truncated to the time that maximizes Equation~\ref{eq:snr-theory}, 
the theoretical SnR of the un-windowed spectra. 
In all panels, the metrics from the un-windowed spectra, i.e., Kaiser $\pi\alpha=0$, is plotted as gray bars towards the left axis. 
The metrics of the rest 4 windows are plotted line point series towards the right axis, 
on the scale of the enhancement percentage. The dashed line marks the 0~\% enhancement level, 
i.e., the value of the metric is equal to the case of the un-windowed spectra.
\label{fig:ocs-summary}}
\end{figure}

\subsection{Preservation of frequency resolution on partially resolved spectra}

In some scenarios, the frequency resolution is more important than the SnR. 
With the wide tunability of the Voigt-1D window, it is able to produce spectra with small FWHMs. 
To evaluate to which degree can the Voigt-1D window preserve the frequency resolution, 
we measured the $J_{10\leftarrow 9}, K=9$ line of \ce{CH3CN} at 183676~MHz, exhibiting a partially resolved hyperfine structure. 
The hyperfine splitting is $\sim 0.8$~MHz, and the Doppler-broadening limit of \ce{CH3CN} is $0.36$~MHz at room temperature.
Therefore, the hyperfine splitting is marginally resolvable under the Doppler-broadening limit. 

Figure~\ref{fig:CH3CN-FID-hfs-256} shows the FID signal of a 60~MHz chirp around the $K=9$ line of pure \ce{CH3CN} vapor measured at 4~$\upmu$bar.
The DAQ delay $t_\text{d}$ was set to 256~ns.
According to the CDMS catalog\cite{CDMSCatalog}, 
the hyperfine splitting of this line has 3 strong components at 183675.955 ($F_{10\leftarrow9}$), 183676.690 ($F_{11\leftarrow10}$), 
and 183676.787 MHz ($F_{9\leftarrow8}$), with relative intensity of 0.997, 1.103, and 0.901.
Since the $F_{11\leftarrow10}$ and $F_{9\leftarrow8}$ components are spaced less than 0.1~MHz, they are unresolvable due to Doppler-broadening.
Therefore, we may simplify the splitting structure into two frequency components which differ by 0.7835~MHz and roughly have a intensity ratio of 1:2.
In Figure~\ref{fig:CH3CN-FID-hfs-256}, the beating of the two frequency components is visible at a period of 1/0.7835~MHz $\approx1.28$~$\upmu$s.
We can directly fit this FID as a product of the decay envelope and the beating of two sine waves (see supplementary material, Section~3.4). 
We fix $a_0=0.4508$~MHz$^2$, and retrieved $b_0=1.005(16)$~MHz.

\begin{figure}[htp!]
\includegraphics{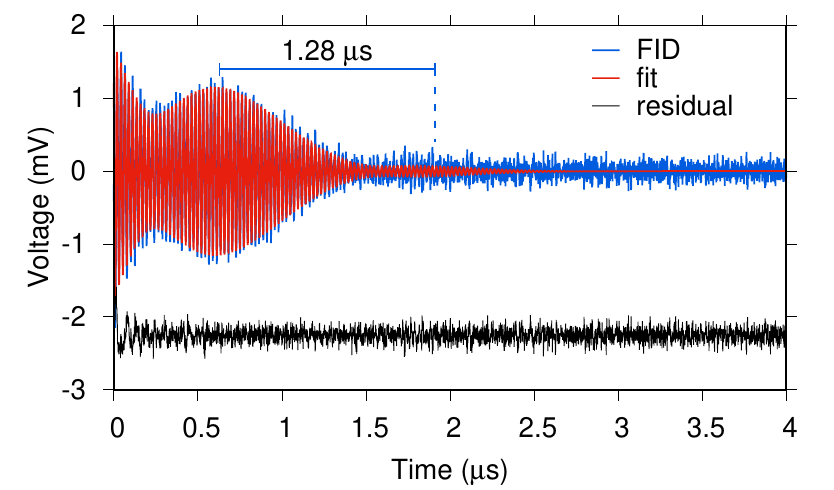}
\caption{The FID of the $J_{10\leftarrow 9}, K=9$ line of \ce{CH3CN} at 183676~MHz, fitted to $f(t) = \exp(-0.4508 t^2 - 1.005(16) t)$. 
The DAQ delay is 256~ns. The LO frequency of this FID is at 183720~MHz, and the chirp bandwidth is 60~MHz. 
The blue trace shows the FID signal,  the red trace shows the least-squared fit, and the black trace shows the fit residual (shifted for clarity). 
\label{fig:CH3CN-FID-hfs-256}}
\end{figure}

Although the beating envelope is apparent in the time domain, the un-windowed spectrum in the frequency domain completely squeezes the two frequency components into a single peak (Figure~\ref{fig:CH3CN-fit}, top panel).
To resolve the two components, frequency resolution, instead of SnR, is the primary goal. 
To begin, we first adjusted the truncation length (3.3~$\upmu$s) of the Kaiser window ($\pi\alpha=8$) so that the two peaks are just about to separate. 
Then, we calculated the Voigt-1D window parameters for high resolution, $a=a_0=0.4508$ and $b=-2\sqrt{a_0}=-1.3428$. 
The fit results of the windowed spectra are also presented in Figure~\ref{fig:CH3CN-fit}.
Both windows are able to resolve the two hyperfine components. 
In the presented results, the Kaiser window forms slightly wider FWHM and also higher SnR than the Voigt-1D window. 
Since the resolution power can be continuously tuned by adjusting the window parameters, 
the Voigt-1D window may be considered having similar performance as the Kaiser window. 

\begin{figure}
\includegraphics{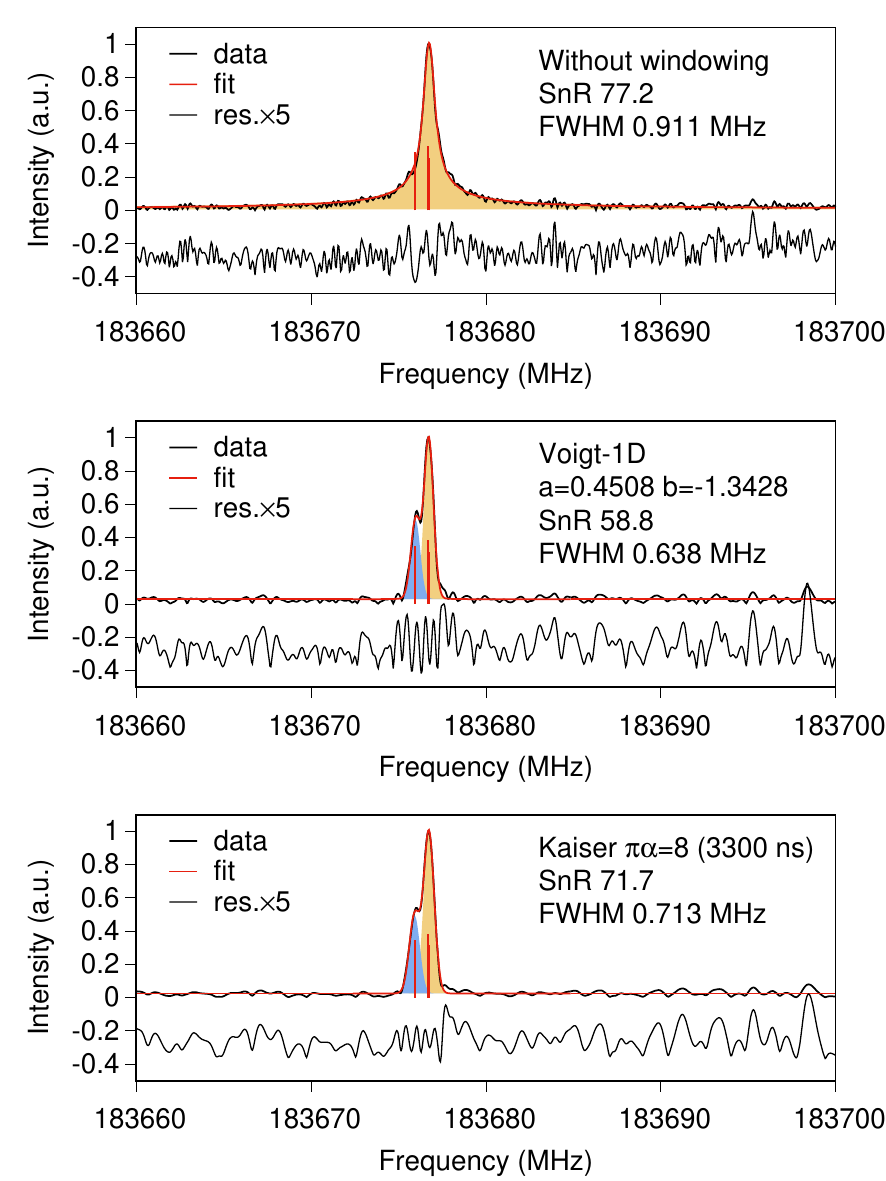}
\caption{Fit results of the $J_{10\leftarrow 9}, K=9$ spectral line of \ce{CH3CN}, treated by the Voigt-1D window and Kaiser window. 
The spectral intensities are normalized to unity to ease the visual comparison of the noise level. 
The black curve shows the spectrum, the red curve shows the overall fit, and the colored areas (blue and yellow) show the individual components of the fit.
The fit residual is plotted below each spectrum.
The red vertical sticks shows the catalog frequency and intensity of the 3 strong hyperfine components.
The noise level is estimated using the standard deviation of the fitted residuals.
\label{fig:CH3CN-fit}}
\end{figure}

It is worth mentioning that, the low frequency beating envelope produced by two closely separated frequency components 
does create a pitfall for parameter selection algorithm of the Voigt-1D windows optimized for highest SnR or SnR/FWHM. 
The parameter selection based on the results in Section~\ref{section:eval-snr}, demonstrated in Figure~\ref{fig:snr-theory}, 
leads to zero Gaussian component in the Window function. 
In this scenario, the window function has a sharp rise at the beginning of the FID, which may coincidentally interfere with the strong beating FID envelope, 
As a result, the spectral lineshape can be altered significantly, presenting asymmetric features.
The readers are referred to Section~4.3 in the supplementary material for a detailed discussion.
Nonetheless, by constraining the $b$ parameter of the Voigt-1D window to $-b_0$, 
so that the majority of the exponential decay in the original FID signal is cancelled out by the window function, 
we can obtain spectral lines with close-to-optimal SnR and correct lineshape.

\subsection{Overall performance on broadband spectrum}

The performance of the Voigt-1D window on room-temperature broadband chirped pulse spectra are tested by numerical simulation, 
because the bandwidth of our DDS-based chirped pulse spectrometer is limited to 0.5~GHz.
Two simulations were performed, one at 2--10~GHz to simulate Lorentzian-dominated microwave chirped pulse spectra, 
and one at 640--650~GHz to simulate Doppler-dominated sub\-millimeter chirped pulse spectra. 
The lines of cis-furfural (\ce{C4H3OCHO}) \cite{Motiyenko2006JMS} were used to simulate the microwave chirp, 
and the lines of N-methylformamide (\ce{HCONHCH3}) \cite{Belloche2017AA} were used to simulate the submillimeter chirp. 
A Lorentzian FWHM of 0.1~MHz was used universally for both simulations. 
More details of the simulation can be found in supplementary material, Section~3.6.
In this simulation, we compared the results of the two Voigt-1D window parameter sets, and the result of the Kaiser window ($\pi\alpha=8$). 

Figure~\ref{fig:broad-cp-mw} shows the simulation of the \ce{C4H3OCHO} spectrum, and 
Figure~\ref{fig:broad-cp-submm} shows the simulation of the \ce{HCONHCH3} spectrum. 
In both simulations, the SnR enhancement of the Voigt-1D window behaves as expected. 
The figures show that the noise level of the un-windowed spectra is similar to the Kaiser windowed spectra, 
and is slightly higher than the spectra from Voigt-1D set (2) window. 
Voigt-1D set (1) window produces spectra with noticeable lower noise which can be observed by eye.
The dashed square box in Figure~\ref{fig:broad-cp-mw} also highlights a weaker line next to a stronger line that
is visually more distinct in the Voigt-1D set (1) windowed spectrum than other spectrum.
The lineshape of the un-windowed magnitude spectra is corrected by the Voigt-1D window, as well as the Kaiser window. 
Overall, the peak resolution of the Voigt-1D set (2) window is similar to the Kaiser window because of their similar FWHM. 
In the right bottom panel of Figure~\ref{fig:broad-cp-mw}, a case shows that Voigt-1D set (2) window performs better in 
resolving a closely spaced doublet than the Kaiser window, because it produces less Gaussian component in the Voigt profile. 
On the other hand, Kaiser window produces best baseline resolution, which is especially useful in identifying lines in the congested regions 
marked out by dashed square boxes in Figure~\ref{fig:broad-cp-submm}.

\begin{figure*}
\centering
\includegraphics[width=\textwidth]{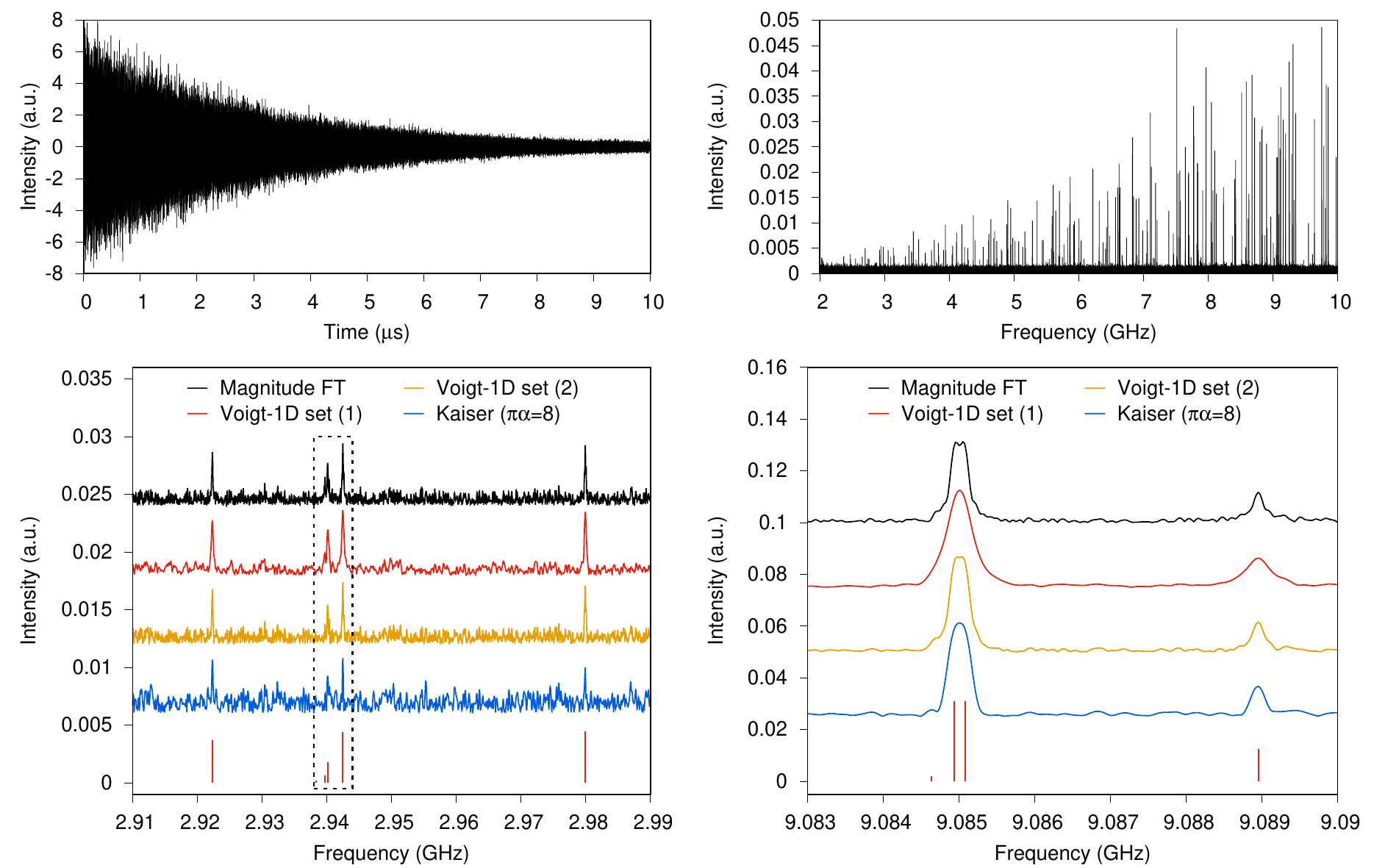}
\caption{Simulated broadband chirped pulse spectrum of \ce{C4H3OCHO} between 2--10 GHz.
The top left panel shows the FID, the top right panel shows the overall spectrum, and the bottom panels show zoomed-in regions.
Red sticks in the bottom panels show the expected transitions from the line catalog. 
The maximum line intensity is normalized to unity for all traces so that the SnR can be directly compared by measuring the noise only. 
Voigt-1D set (1) uses $(a, b)=(0, 0.9461\text{ MHz})$ for maximizing SnR, and Voigt-1D set (2) uses $(a, b)=(0, \text{ 0.3170})$ for maximizing SnR/FWHM.
\label{fig:broad-cp-mw}}
\end{figure*}

\begin{figure*}
\centering
\includegraphics[width=\textwidth]{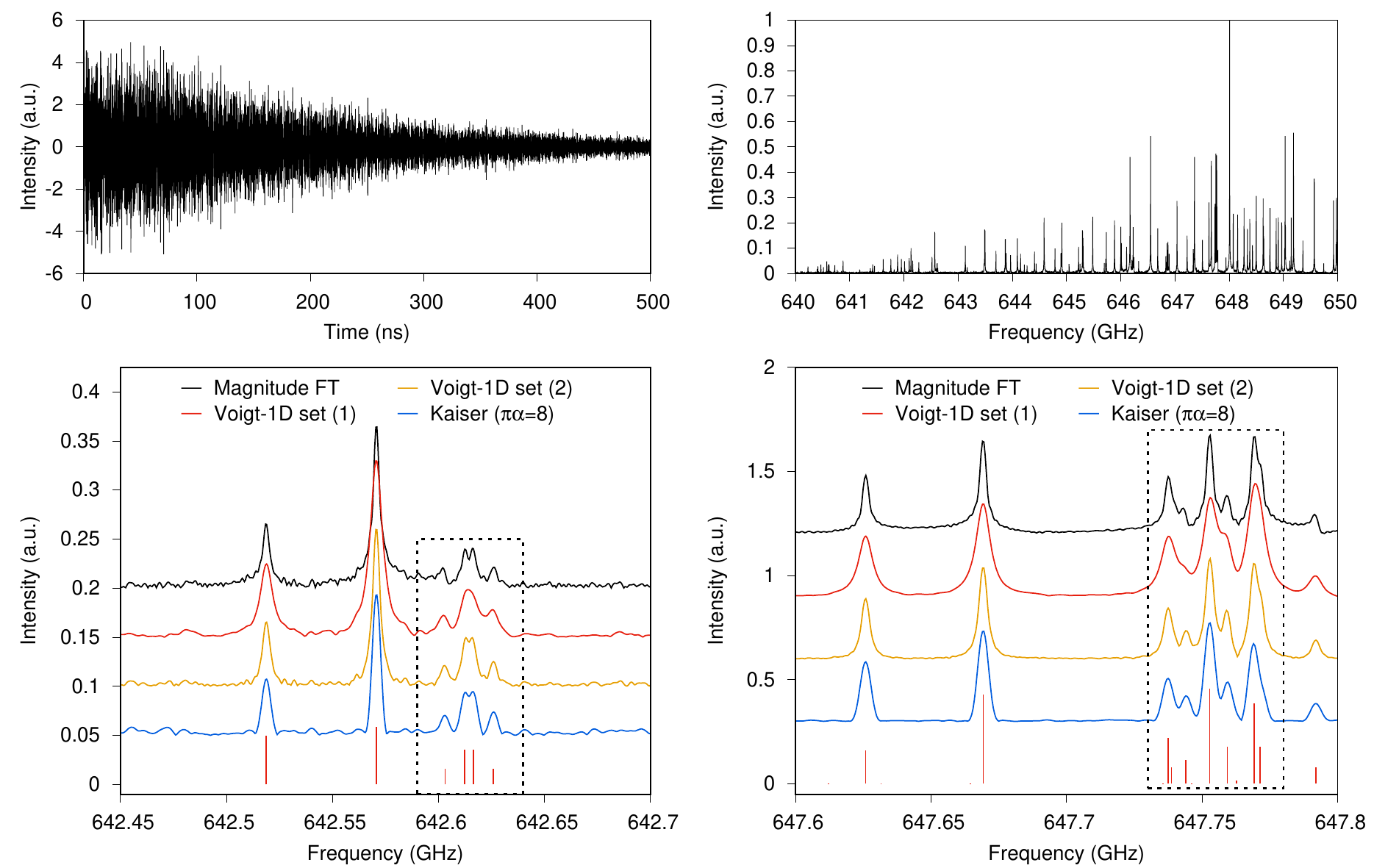}
\caption{Simulated broadband chirped pulse spectrum of \ce{HCONHCH3} at 640--650 GHz.
The top left panel shows the FID, the top right panel shows the overall spectrum, and the bottom panels show zoomed-in regions.
Red sticks in the bottom panels show the expected transitions from the line catalog. 
The maximum line intensity is normalized to 1 for all traces so that the SnR can be directly compared by measuring the noise only. 
Voigt-1D set (1) uses $(a, b)=(0, 15.7789\text{ MHz})$ for maximizing SnR, and Voigt-1D set (2) uses $(a, b)=(0, 6.5515\text{ MHz})$ for maximizing SnR/FWHM.
\label{fig:broad-cp-submm}}
\end{figure*}

\section{Conclusion}

In conclusion, we have proposed a specific window function, noted as the ``Voigt-1D'' window, 
for treating the data of chirped pulse experiments. 
The window function takes the form $w(t; a, b) = t\exp(-at^2-bt)/M$, where $t$ is the time variable, 
$a$ and $b$ are two adjustable parameters, and $M$ is the normalization factor.
The window function corrects the spectral lineshape, offers wide tunability by adjusting its two parameters,
suppresses baseline ripples from zero-padding, 
and is able to generate high SnR spectra with properly chosen parameters. 
We have derived the mathematical SnR equations for a chirped pulse spectral line treated by the Voigt-1D window, 
and discussed the guideline of parameter selection based on these mathematical expressions.
The programmable routines to find two optimal parameter sets, one for achieving maximum SnR, 
and one for achieving maximum SnR/FWHM, are proposed.

Performance of the Voigt-1D window is evaluated by treating real chirped pulse experimental data of OCS and \ce{CH3CN} lines,
and by treating simulated broadband chirped pulse data. 
The results are compared with un-windowed magnitude Fourier transform spectra, 
and with results treated by Kaiser-Bessel windows, which are commonly used in current chirped pulse literature.
Our results show that the Voigt-1D window with maximum SnR/FWHM is able to enhance the SnR by 100~\% on average, and sometimes up to 250~\%
than the un-windowed magnitude spectra, at a cost of only 25~\% wider FWHM. 
The Voigt-1D window with maximum SnR can achieve slightly higher SnR enhancement at the cost of almost doubling the FWHM of the un-windowed spectra. 
Compared with Kaiser window ($\pi\alpha=4$), the SnR enhancement of the Voigt-1D window is higher, whereas the FWHM broadening is lower. 
When resolving closely spaced lines with similar intensities, 
the Voigt-1D is able to reach similar, and sometimes better, performance to the Kaiser window ($\pi\alpha=8$). 
When baseline resolution is the primary goal, such as decomposing congested spectra and identifying weak satellite lines on the wings of strong lines, 
Kaiser window is still the preferred choice. 

\section{Supplementary material}

Supplementary material includes additional mathematics of the Voigt-1D window, experimental, simulation, and data fitting details, as well as some supplementary results such as the performance of “averaged” window parameters on broadband chirp. The supporting Python code for the optimization of Voigt-1D window function parameters is available at https://doi.org/10.5281/zenodo.4555759 and from the authors upon reasonable request.

\section{Acknowledgement}

The authors thank the anonymous referees for helpful comments that improved the quality and robustness of this manuscript. 
This work was funded by the French ANR Labex CaPPA through the PIA under Contract No.\ ANR-11-LABX-0005-01, 
the Regional Council of Hauts-de-France, the European Funds for Regional Economic Development (FEDER), 
and the contract CPER CLIMIBIO. 
L. Zou thanks the financial support from the European Union's Horizon 2020 research and innovation programme 
under the Marie Sk{\l}odowska-Curie Individual Fellowship grant No.\ 894508.

\section{Contributor Statement}

\textbf{Luyao Zou}: Methodology, Investigation, Formal analysis, Visualization, Writing - Original Draft, Writing - Review \& Editing.
\textbf{Roman Motiyenko}: Conceptualization, Methodology, Writing - Review \& Editing, Supervision.

\bibliographystyle{elsarticle-num}

\end{document}

\maketitle

\section{The magnitude Fourier transform (FT) of the FID\label{section:magnitude-FT}}

In this section, we demonstrate the deviation of spectral lineshape from Voigt profile 
for the magnitude FT of an FID signal.
We then prove that the Voigt-1D corrects the lineshape back to Voigt profile. 

To start, consider the envelope of a pure exponential FID: $f(t) = e^{-bt}, t\geq 0$.
It can be also viewed as $f(t) = u(t)e^{-|b|t}$: 
the product of the two-sided exponential decay $e^{-b|t|}$, which is an even function, and the Heaviside unit step function $u(t)$.

The FT of $f(t)$ is 
\begin{equation}
\hat{f}(x) = \frac{1}{b + 2\pi i x} \equiv \frac{b - 2\pi i x}{b^2 + 4\pi^2 x^2}
\end{equation}
where $i$ is the imaginary unit. Its magnitude is 
\begin{equation}
|\hat{f}(x)| = \frac{1}{\sqrt{b^2 + 4\pi^2 x^2}}
\end{equation}
It shows that the real part (absorption) of $\hat{f}(x)$ is a Lorentzian profile, 
whereas the magnitude of $\hat{f}(x)$, $|\hat{f}(x)|$, is the square root of a Lorentzian profile, 
and therefore has wider wings than a Lorentzian. 

The magnitude of the first derivative of $\hat{f}(x)$ is again a Lorentzian profile.
\begin{equation}
\frac{\mathrm{d}}{\mathrm{d} x} \hat{f}(x) = -\frac{8\pi^2 b x}{(b^2 + 4\pi^2 x^2)^2} - \frac{2\pi i (b^2-4\pi^2x^2)}{(b^2 + 4\pi^2 x^2)^2}
\end{equation}
\begin{equation}
\begin{aligned}
\Big|\frac{\mathrm{d}}{\mathrm{d} x} \hat{f}(x) \Big|^2 &=
  \frac{1}{(b^2 + 4\pi^2 x^2)^4}\Big(64\pi^4 b^2x^2 + 4\pi^2((b^2 - 4\pi^2 x^2)^2)\Big) \\
  &= \frac{1}{(b^2 + 4\pi^2 x^2)^4}\Big(64\pi^4 b^2x^2 + 4\pi^2\big(b^4 + 16\pi^4 x^4 - 8\pi^2 b^2x^2\big)\Big) \\
  &= \frac{1}{(b^2 + 4\pi^2 x^2)^4}\Big(4\pi^2(b^4 + 16\pi^4x^4 + 8\pi^2b^2x^2))\Big) \\
  &= \frac{1}{(b^2 + 4\pi^2 x^2)^4}\Big(4\pi^2(b^2 + 4\pi^2x^2)^2\Big) \\
 \Big|\frac{\mathrm{d}}{\mathrm{d} x} \hat{f}(x) \Big| &= \frac{2\pi (b^2 + 4\pi^2x^2)}{(b^2 + 4\pi^2 x^2)^2} = 
	\frac{2\pi}{b^2 + 4\pi^2x^2} \quad \text{(pure Lorentzian)}
\end{aligned}
\end{equation}
According to the FT property
\begin{equation}
\widehat{tf(t)}(x) = \frac{i}{2\pi}\frac{\mathrm{d}}{\mathrm{d} x}\hat{f}(x)
\end{equation}
we see that the magnitude FT of $tf(t)$ is also a pure Lorentzian profile.
More importantly, this Lorentzian profile has the same FWHM as the original one $\mathfrak{Re}[\hat{f}(x)]$.

When the FID has both exponential and Gaussian decay, 
the FID envelope can be rewritten as the product of a pure exponential decay and a Gaussian decay, i.e., 
\begin{equation}
F(t) = f(t) g(t) = u(t) e^{-b|t|} e^{-at^2}
\end{equation}
According to the convolution theorem, the FT of $F(t)$ is the convolution of the FT of $f(t)$ and $g(t)$:
$\hat{F}(x) = \hat{f}(x) (*) \hat{g}(x)$.
Also note the property of the derivative of convolution:
if $h(x) = f(x) (*) g(x)$, then $h'(x) = f'(x) (*) g(x) = f(x) (*) g'(x)$.
The magnitude FT of $tF(t) = tu(t)e^{-bt} e^{-at^2}$ is therefore equal to 
\begin{equation}
\big|\widehat{tF(t)}\big| = \Big|\frac{i}{2\pi}\hat{F}'(x)\Big| = \frac{1}{2\pi}|\hat{F}'(x)|
\end{equation}
Since $\hat{g}(x)$ is the FT of a Gaussian ($t\in(-\infty, +\infty)$), $\hat{g}(x)$ is also a Gaussian and pure real.
As a result, 
\begin{equation}
\big|\hat{F}'(x)\big| = \big|\hat{f}'(x)\big| (*) \hat{g}(x)
\end{equation}
which is a Voigt profile, and so is $\big|\widehat{tF(t)}\big| $.

\section{The spectral density of white Gaussian noise without and with windowing\label{section:magnitude-FT-noise}}

To calculate the frequency domain noise, we first consider the signal as continuous, and then generalize the result to the discrete signal.

Note the time domain white Gaussian noise as a Gaussian random process $\{X_t\}$ with mean $\mu_t=0$ and variance $\sigma$ for all $t\in[0, T]$.
The auto-correlation function of $\{X_t\}$ is $R_{XX}(\tau)=\sigma^2 \delta(\tau)$.
According to the Wiener-Khinchin theorem, the power spectral density (PSD) of $\{X_t\}$ is the Fourier transform of $R_{XX}(\tau)$:
\begin{equation}
S_{XX}(\nu) = \int_{-T}^{T} R_{XX}(\tau) e^{-2\pi i \nu\tau}\mathrm{d}\tau
= \sigma^2 \int_{-T}^{T} \delta(\tau) e^{-2\pi i \nu\tau}\mathrm{d}\tau = \sigma^2
\end{equation}
Now, in the discrete case, the variance of each frequency domain data point is then the PSD divided by the frequency resolution $\Delta\nu$, 
which is equal to $1/T$. 
Therefore, the frequency domain noise is $\sqrt{TS_{XX}(\nu)} = \sqrt{T}\sigma$.

When the noise is multiplied by a window function as $\{Y_t\} = \{X_tw(t)\}$, $t\in[0, T]$,
its auto-correlation function becomes
\begin{equation}
\begin{aligned}
R_{YY}(\tau) & = \big\langle Y_{t} Y_{t+\tau} \big\rangle = \big\langle X_{t} X_{t+\tau} w(t)w(t+\tau) \big\rangle \\
& = \frac{\sigma^2 \delta(\tau)}{T} \int_{0}^{T}w(t) w(t+\tau)\delta(\tau)\mathrm{d}t 
= \frac{\sigma^2 \delta(\tau)}{T} \int_{0}^{T}w^2(t) \mathrm{d}t 
\end{aligned}
\end{equation}
Therefore, its PSD is 
\begin{equation}
S_{YY}(\nu) = \frac{\sigma^2}{T} \int_{0}^{T}w^2(t) \mathrm{d}t 
\end{equation}
The spectral noise is then 
\begin{equation}
\text{noise} = \sqrt{T S_{YY}(\nu)} = \sigma \sqrt{\int_{0}^{T}w^2(t) \mathrm{d}t }
\end{equation}
We may see that if $w(t)\equiv 1$, the noise goes back to $\sqrt{T}\sigma$.

\section{Experimental, Simulation, and Data Fitting Details\label{section:exp-sim-fitting}}

To evaluate the performance of the Voigt-1D window function, we performed both numerical simulation and treatment on real experimental data. 
Both simulation and the treatment of experimental data were performed using Python (version 3.6) with its \verb|numpy| and \verb|scipy| modules. 
The least-square fits were performed using the \verb|lmfit| module, which is based on \verb|scipy|.

\subsection{Experimental\label{section:experimental}}

The experiment was conducted using the (sub)millimeter chirped pulse spectrometer whose details are described elsewhere \cite{Zou2020RSI}.
In brief, the chirp signal ($<20$~MHz bandwidth) was generated using a direct digital synthesizer (AD9915), 
filtered, and up-converted to 12--18~GHz to pump a (sub)millimeter multiplier chain system (Virginia Diodes, Inc.).
After frequency multiplication, the chirp bandwidth becomes 60--300~MHz. 
A corresponding zero-biased detector was used as a harmonic mixer to detect the chirp signal, 
which was then amplified and digitized by a digitizer card (GaGe RazorMax, 16-bit, 1 Gs/s).
Since in this study, the focus is on the window function itself,
the choice of the molecule and the exact experimental conditions is somewhat arbitrary. 
We used carbonyl sulfide (OCS) to evaluate the window function performance on FIDs containing a single line, 
and acetonitrile (\ce{CH3CN}) to evaluate the window function performance on closely spaced hyperfine lines. 

The experiment measurements of the OCS lines we used in this article are those reported in \citet{Zou2020RSI}.
In those experiments, the lines were measured using a ``dual direction mode'' that alternates the frequency sweep direction. 
In this paper, however, we separated the FID data of the two directions, and treated them separately, 
so that the FID signal and the spectral lineshape in the frequency domain can both be unambiguously modeled.

The experiment measurement of the \ce{CH3CN} lines was conducted at room temperature in a cell of 1~m long. 
The cell was designed for pressure-broadening experiment and its pressure was monitored by a MKS 627A baratron. 
\ce{CH3CN} (Sigma Aldrich) vapor was first introduced to the cell without additional treatment, 
and then slowly pumped out to reach the desired pressure of 4~$\upmu$Bar.
The spectra were measured using the WR5.1 multiplier and detector (140--220~GHz).
The chirp excitation lasts 512~ns, sweeping from 183660~MHz to 183720~MHz,  
and we wait additionally 256~ns after the chirp ends before starting DAQ. 
The local oscillator frequency of the heterodyne detection was 183720~MHz.

\subsection{Numerical Simulation}

The equations in Section~2.2 in the main article were first validated by numerical simulation of the discrete Fourier transform of FID with Gaussian noise. 
Using these equations, we scanned the SnR values over a grid of $a$, $b$ for a few selected $a_0$, $b_0$ combinations to obtain Figure~2 in the main article.
Then we treated the experimental data based on the same computer code. 
In the end, we simulated broadband chirped pulse spectra that contain hundreds of lines,
and evaluated the performance of the Voigt-1D window on the simulated spectrum.

\subsection{Maximizing theoretical SnR and SnR/FWHM\label{section:theoretical-snr}}

The $(a,b)$ values for highest SnR and SnR/FWHM were calculated numerically. 
For the highest SnR, we defined the model as the negation of $s(a_0, b_0, a, b)$ (Eq.~12 in the main article), 
and use the default least-square minimizer in \verb|lmfit| (Levenberg-Marquardt method) to find the minimum point $(a, b)$. 
The Jacobian was estimated internally by the algorithm. 

For SnR/FWHM, we first defined a numerical function to find the FWHM $\gamma_V(a, b)$, instead of using approximation equations\cite{Olivero1977JQSRT}. 
The analytical expression of the magnitude FT of the window function is 
\begin{equation}
\begin{aligned}
|y(\omega)| &= \bigg| \int_0^{+\infty} w(t; a, b)\exp(-i\omega t)\mathrm{d} t \bigg|\\
&= \frac{1}{2a}\bigg| 1 + i \sqrt{\pi} \exp(\frac{(b+i\omega)^2}{4a}) \text{erfc}(\frac{b + i\omega}{2\sqrt{a}}) \bigg| \\ 
&= \frac{1}{2a}\bigg| 1 + i \sqrt{\pi} z\text{wofz}(z)\bigg|
\end{aligned}
\end{equation}
where we use angular frequency $\omega = 2\pi x$ to simplify the equation.
$\text{wofz}(z)$ is the Faddeeva function defined as $\text{wofz}(z) := e^{-z^2}\text{erfc}(-iz)$, where $z$ is defined as
\begin{equation}
z = \frac{-\omega + ib}{2\sqrt{a}}
\end{equation}

The FWHM of $|y(\omega)|$ is found numerical by solving the equation of $|y(\text{FWHM}/2)| = |y(0)|/2$.
We use the \verb|scipy.optimize.root| function to find the root of $|y(\text{FWHM}/2)| = |y(0)|/2$.
After finding the FWHM, the model function $-s(a_0, b_0, a, b)/\gamma_V(a, b)$ was minimized to find the $(a, b)$ value for highest SnR/FWHM.

\subsection{FID fit\label{section:FID-fit}}

The FID fit was performed using the \verb|lmfit| module. 

For OCS lines, the FID only contains one frequency component. 
There are several ways to construct the model equation, depending on how the DAQ delay time $t_0$ is expressed. 
Ideally, one would write 
\begin{equation}
f(t) = s \exp(-a_0 (t+t_0)^2 - b_0(t+t_0))\sin(2\pi f_\text{if} (t+t_0) + \phi) + p_0
\end{equation}
where $a_0$ and $b_0$ are the initial parameters that describes the FID envelope, 
$f_\text{if}=|f_\text{lo} - f_\text{s}|$ is the intermediate frequency, $f_\text{s}$ is the line frequency and $f_\text{lo}$ is the local oscillator frequency,
$\phi$ is the initial phase, $s$ is the intensity scalar, and $p_0$ is the baseline drift. 
To avoid numerical instability, we simplified the above equation to 
\begin{equation}
f(t) = s \exp(-a_0t^2-b_0t)\sin(2\pi f_\text{if} t + \phi) + p_0
\end{equation}
so that $t_0$ is merged into the effective intensity $s$ and phase $\phi$ terms. 
For each FID record, 
before the least-square fit, the data was first filtered by a 4\textsuperscript{th} order Chebyshev type II high-pass filter, 
with minimum attenuation of 60~dB at the stop band and critical frequency of 10~MHz. 
The filter was generated using the \verb|scipy.signal.cheby2| function.
The filter was applied to the FID data twice, forward and backward so that the phase is not altered. 
The initial guesses were set as $b_0=2a_0t_0$, $s=\text{max}\{f(t)\}$, $p_0=0$, and $\phi$ selected manually.
During the fit, $a_0$ was fixed to $\pi^2\gamma_G^2/(4\ln2)$, where $\gamma_G$ is the Doppler FWHM,  and other parameters were fitted. 

For the \ce{CH3CN} hyperfine line, we treat the FID as two beating frequencies.
We may write
\begin{equation}
f(t) = p_0 + \exp(-a_0t^2-b_0t) \Big( s_1 \sin(2\pi f_1 (t+t_1) + \phi_1) + s_2 \sin(2\pi f_2 (t+t_2) + \phi_2) \Big )
\end{equation}
Here, the meaning of $p_0$, $a_0$, and $b_0$ are the same as the fit model for OCS lines. 
$s_i$, $f_i$, $t_i$, and $\phi_i$  ($i=1, 2$) are the intensity scalar, intermediate frequency, DAQ delay time, and arbitrary initial phase 
for the $i$\textsuperscript{th} frequency components.
Again, for numerical stability, we simplified the above equation to 
\begin{equation}
f(t) = p_0 + \exp(-a_0t^2-b_0t) \Big( s_1 \sin(2\pi f_1 t + \phi_1) + s_2 \sin(2\pi f_2 t + \phi_2) \Big )
\end{equation}
Before the least-square fit, the data was first filtered by a 3\textsuperscript{rd} order Butterworth high-pass filter, 
with a critical frequency of 15~MHz. 
The filter was generated using the \verb|scipy.signal.butter| function.
The filter was applied to the FID data twice, forward and backward so that the phase is not altered. 
In the fit, $a_0$ was fixed to $\pi^2\gamma_G^2/(4\ln2)$, where $\gamma_G$ is the Doppler FWHM using the average frequency of the two frequencies. 
The ratio $s_1/s_2$ was fixed to $2:1$.
$p_0$, $b_0$, $f_i$, $\phi_i$ and $s_i$ were fitted.

\subsection{Spectral fit\label{section:spectral-fit}}

The spectral fit was also performed using the \verb|lmfit| module. 

The frequency domain Voigt profile was defined as 
\begin{equation}
v(x; x_0, \gamma_G, \gamma_L) = \mathfrak{Re}\bigg[\text{wofz}\Big(\frac{x + i \gamma_L / 2}{\sqrt{2}\sigma}\Big)\bigg], \quad \sigma = \frac{\gamma_G}{2\sqrt{2\ln 2}}
\label{eq:voigt-profile}
\end{equation}
where $\gamma_G$, $\gamma_L$ are the FWHM of Gaussian and Lorentzian components, 
and wofz($z$) is the Faddeeva function, provided by the \verb|scipy.special.wofz| function. 
The FWHM of the line was calculated using fitted $\gamma_G$ and $\gamma_L$ values, and the approximation 
$\gamma_V =  0.5346 \gamma_L + \sqrt{0.2166 \gamma_L^2 + \gamma_G^2}$\cite{Olivero1977JQSRT}.

The frequency domain complex Voigt profile $vc(x)$ was defined using the identical Faddeeva function as in Equation~\ref{eq:voigt-profile},
by replacing the $\mathfrak{Re}$ operation by taking the magnitude of the function. 
The FWHM of the complex Voigt profile was determined numerically, 
by finding the root of $vc(\text{FWHM}/2) = vc(0)/2$.

For numerical stability, the least-square fit was performed under a normalized condition. 
Before the fit, the $x$ array, so as the associated $x_0$, $\gamma_G$, and $\gamma_L$ values, was rescaled to $[-1, 1]$, 
and the maximum of the $y$ array was normalized to 1. 
The fit result was then converted back to their original scale. 
We took the standard deviation of the residual as the noise for SnR calculation, 
and therefore imperfect fit of the line profile causes an effective decrease of the SnR.

The fit of a pure Gaussian profile was conducted in the similar manner, using a pure Gaussian function as the fitting model. 

For OCS lines, the noise was measured as the standard deviation of the fit residual within $\pm20$~MHz of the line center. 
For \ce{CH3CN} lines, the noise was measured as the standard deviation of the fit residual from 183660 to 183700 MHz.

\subsection{Simulation of broadband CP\label{section:sim-broadband}}

For broadband CP signal, we simulated a 500~ns pulse, followed by a $t_\text{d}=250$~ns delay before DAQ.
For the 2--10~GHz chirp, the frequency was directly simulated in the time domain.
For the 640--650~GHz chirp, the local oscillator frequency was set at 640~GHz, 
and therefore the intermediate frequency of 0--10~GHz was simulated in the time domain. 
The sampling frequency was set to 50~GHz, which is beyond the Nyquist frequency of the 0--10~GHz signal for better sampling. 
The peak frequencies $f_i$ and peak intensities $s_i$ were retrieved from the SPFIT catalog files generated by the fit of rotational lines of the corresponding molecules. 
The initial phases $\phi_i$ were randomly generated using a uniform distribution in the range of 0--$2\pi$ rad. 
For each peak, the Doppler FWHM was calculated at 300~K. 
For each peak, a unique delay time $t_i$ was calculated based on its exact excitation time, using the following equation: 
\begin{equation}
t_i = \frac{f_\text{up} - f_i}{f_\text{bw}}t_\text{cp} + t_\text{d}
\end{equation}
where $f_\text{bw}$ is the chirp bandwidth (10~GHz for both simulations),
$f_\text{up}$ is the maximum frequency of the chirp, and $t_\text{cp}$ is the chirp duration.
The sum of FID signals from all peaks, $\sum_i s_i \exp(-a_0 (t+t_i)^2 - b_0(t+t_i)) \cos(2\pi f_i (t+t_i) + \phi_i)$, 
was added to a white Gaussian noise, created by a normal distribution random generator. 
The final signal was saved and used to generate the frequency domain spectrum. 
When calculating the magnitude FT spectrum, a zero-padding factor of 3 was added to the FID signal for smoother spectral lines.

\section{Supplementary Results}

\subsection{The performance of ``averaged'' window parameters on broadband chirp}

Following the discussion in Section 2.5 and 3.1 of the main article, 
We test how the ``averaged'' parameters of the Voigt-1D window function will perform over the broadband chirped pulse spectra. 
For each molecular line at frequency $f_i$, the actual FID initial parameters are 
\begin{align}
a_0 & = \frac{\pi^2\gamma_G^2}{4\ln2} = \frac{\pi^2(\delta g)^2f_i^2}{4\ln2} \\
b_0 & = \pi\gamma_L + 2 a_0 t_i = \pi\gamma_L + 2 a_0 \Big( \frac{f_\text{up} - f_i}{f_\text{bw}}t_\text{cp} + t_\text{d} \Big)
\end{align}
where $\delta g$ is the Doppler broadening coefficient, and other labels follow the definition in Section 3.6.

Based on experimental practices, we may find 3 characteristic scenarios:
\begin{enumerate}
\item Microwave, Doppler dominated. This usually occurs in jet-cooled conditions, where the Doppler effect from the residual velocity dispersion of the jet is the main dephasing mechanism. 
Because the Doppler width is proportional to the line frequency, and because in microwave chirps, the ratio of starting and ending frequency is large (i.e., a factor of 4 for 2--8~GHz chirp),
the change of $a_0$ and $b_0$ parameters for each line is also significant across the chirp band. 
\item (Sub)millimeter, Doppler dominated. This usually occurs in room-temperature, low pressure measurements, where the thermodynamic Doppler broadening
is much larger than the pressure broadening, and therefore is the dominant dephasing mechanism. 
The chirp bandwidth is usually small compared to the working frequency range, and therefore the percentage change of $a_0$ and $b_0$ parameters for each line 
is much less than that in scenario 1.
\item Lorentzian dominated. This occurs when the pressure broadening is the dominant dephasing mechanism, and therefore the dominant parameter of the FID envelope shape. 
\end{enumerate}

In scenario 1, we refer to the experimental conditions described in \cite{Fatima2020PCCP}. 
$\gamma_G$ is on the order of 200~kHz, the chirp bandwidth the lowest frequency is 2--8~GHz, and the pressure broadening $\gamma_L\approx 0$.
The chirp excitation $t_\text{cp}$ lasts 4~$\upmu$s, and the DAQ dead time $t_\text{d}$ is not described. 
According to these settings, we may configure $a_0$ to range from 0.036~MHz$^2$ to 0.570~MHz$^2$ ($\gamma_G$ from 100 to 400 kHz), 
and $b_0$ to range from 0.285 to 0~MHz, assuming $t_\text{d}=0$.
This is the most dramatic change of $a_0$ and $b_0$ in all 3 scenarios.

In scenario 2, we refer to the experimental conditions described in \cite{Gerecht2011OptExp}. 
$\gamma_G$ is on the order of 1.8~MHz, the chirp bandwidth is $9.6\approx 10$~GHz, and the working frequency is around 600~GHz. 
The chirp excitation $t_\text{cp}$ lasts 25~ns, and the DAQ dead time $t_\text{d}$ is not described. 
According to these settings, we may configure $a_0$ to range from 11.15~MHz$^2$ to 11.54~MHz$^2$($\gamma_G$ from 1.77 to 1.8 MHz), 
and $b_0$ to range from 0.558 to 0~MHz, assuming $t_\text{d}=0$.

In scenario 3, we may assume $\gamma_L$ of 2~MHz, and $\gamma_G$ of 200~kHz. 
To let the change of $a_0$ again dramatic, we may consider the same chirp settings with scenario 1, and additionally let $t_\text{d}$=1~$\upmu$s. 
According to these settings, we may configure $a_0$ to range from 0.036~MHz$^2$ to 0.569~MHz$^2$ ($\gamma_G$ from 100 to 400 kHz), 
and $b_0$ to range from 6.639 to 7.422~MHz.

In each scenario, 
we calculate the actual $a_0$ and $b_0$ values for a grid of $f_i$ spanning from the full chirp band,
and then find the the optimal Voigt-1D window function parameters $(a_i, b_i)$ for SnR and SnR/FWHM using our proposed procedure in Section 3.1 of the main article.
Then, we calculate the optimal Voigt-1D window function parameters $(a_\text{avg}, b_\text{avg})$ from a fixed $a_{0_\text{avg}}$ and $b_{0_\text{avg}}$, 
which correspond to the $(a_0, b_0)$ of a line at the center frequency of the chirp band. 
We apply both $(a_i, b_i)$ and $(a_\text{avg}, b_\text{avg})$ to each line at $f_i$, 
and we compare the SnR and SnR/FWHM produced by the two different window function parameters, using Equation~13 in the main article.
In theory, the result from $(a_\text{avg}, b_\text{avg})$ should be lower than the result from $(a_i, b_i)$, 
because the latter is the actual optimal window.

\begin{figure}[htp!]
\centering 
\includegraphics[width=\textwidth]{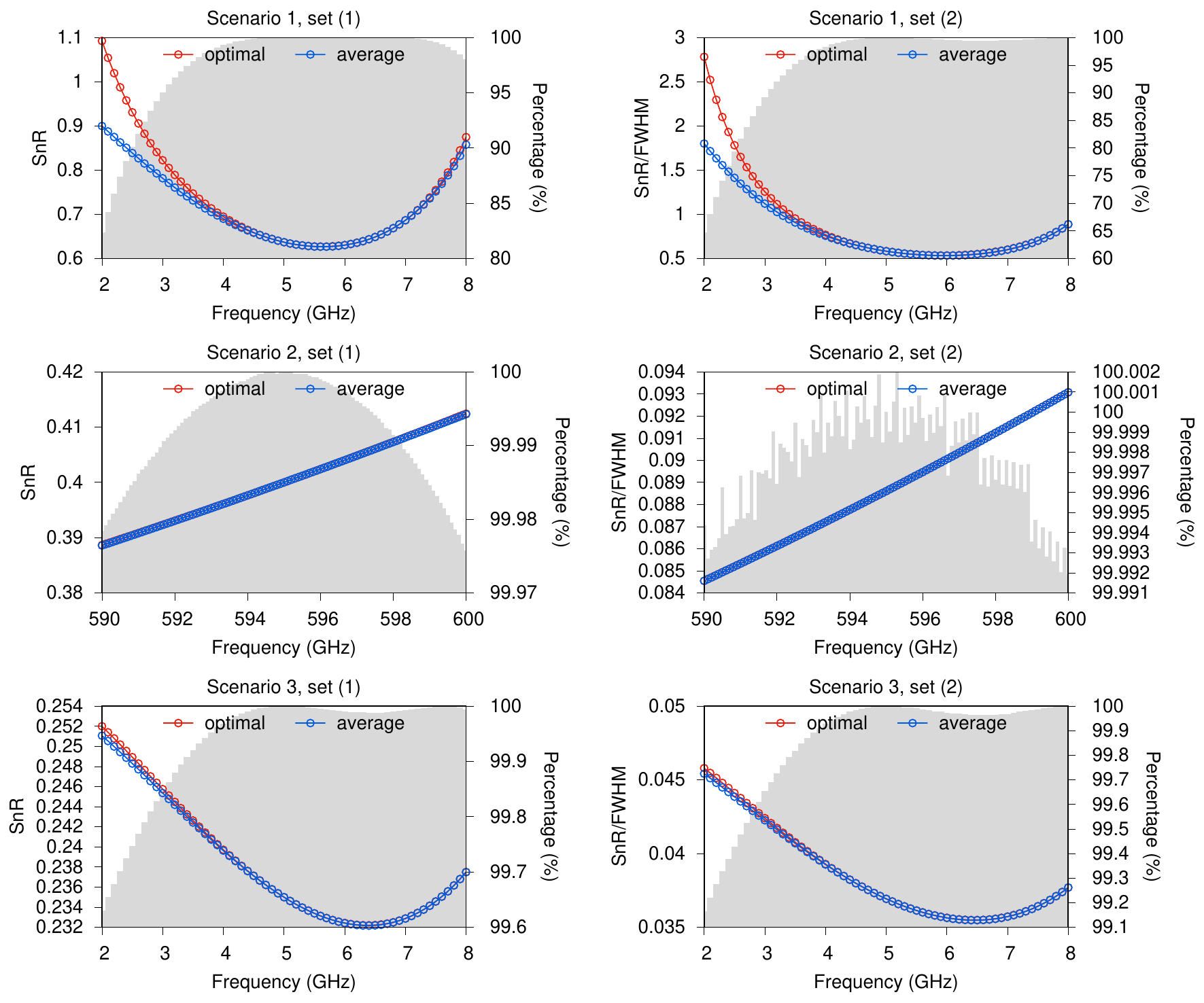}
\caption{Performance comparison of the ``averaged'' window parameters and the optimal window parameters for broadband chirped pulse spectra. 
Each row corresponds to a scenario discussed in Section 4.1; the left column applies the Voigt-1D parameter set (1) that maximizes SnR, 
and the right column applies the Voigt-1D parameter set (2) that maximizes SnR/FWHM. 
\label{fig:v1d-param-avg}}
\end{figure}

Figure~\ref{fig:v1d-param-avg} shows the results for the 3 scenarios. 
In this figure, each row corresponds to a scenario. 
The left column shows Voigt-1D window parameter set (1) that maximizes SnR, 
and the right column shows Voigt-1D window parameter set (2) that maximizes SnR/FWHM. 
The actual optimal results obtained by applying window parameters $(a_i, b_i)$ are plotted in red, 
and the ``averaged'' results obtained by applying window parameters  $(a_\text{avg}, b_\text{avg})$ are plotted in blue. 
The ratio between the ``averaged'' results and the optimal results are plotted as grey boxes on the right axis. 

We can see that for scenario 2 and 3, the results are almost identical across the whole chirp band. 
The ``averaged'' results reach over 99~\% of the actual optimal results.
This is expected because in both scenarios, the percent change of $a_0$ and $b_0$ across the whole chirp band is quite small, 
and therefore the ``averaged'' window parameters are near the optimal point. 

The result of scenario 1 shows larger difference between the ``averaged'' and optimal, 
because $a_0$ changes by a factor of 4. 
However, we can see that the ``averaged'' SnR is over 80~\% of the optimal SnR, 
and the performance drop is most significant at the beginning of the chirp, which is subjected to a long $t_0$ delay. 
The performance soon reaches over 90~\% for frequencies $>2.5$~GHz. 
This means less than 10~\% of the whole chirp band will have slightly higher performance loss by the treatment of the ``averaged'' window, 
and such loss is acceptable.
For SnR/FWHM, the loss is as high as 40~\% (100~\%$-$60~\%). 
This is because the ``averaged'' window produces larger FWHM than the optimal window at the beginning of the chirp.
Same as the SnR results, the performance quickly reaches over 80~\% of the optimal for frequencies $>2.5$~GHz. 
Considering this scenario is the most extreme case, we can conclude that in general, the strategy of using ``averaged'' window parameters
for broadband chirped pulse spectra is reasonable. 

\subsection{The OCS samples and fits}

\begin{figure*}[htp!]
\centering
\includegraphics[width=\textwidth]{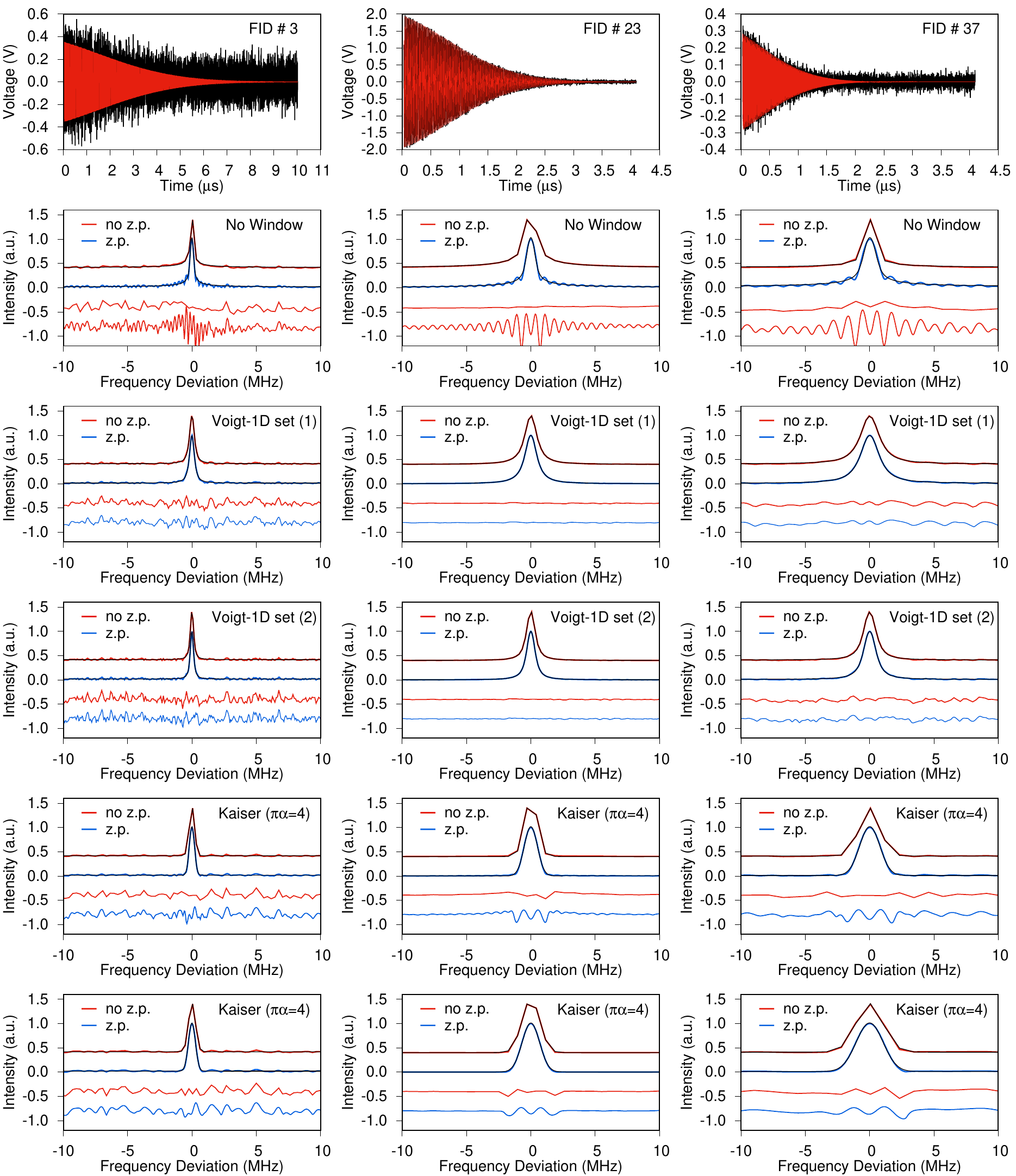}
\caption{The FID and spectra of 3 OCS lines. 
From left column to right column, the transition frequencies are about 61~GHz, 195~GHz, and 260~GHz, respectively.
The top row shows the FID signal (black) along with the least-square fit (red) of the FID. 
The following rows show the frequency domain spectra of the corresponding FID, without or with zero-padding (z.p.) to 20~$\upmu$s.
The spectra are plotted with respect to the peak frequencies. 
The window functions applied, following the row order, are no window (Kaiser $\pi\alpha=0$), Voigt-1D set (1) (for maximum SnR), Voigt-1D set (2) (for maximum SnR/FWHM),
Kaiser $\pi\alpha=4$, and Kaiser $\pi\alpha=8$. 
The fit residuals, $\times$5, are plotted under the spectral lines. 
\label{fig:ocs-samples}}
\end{figure*}

Figure~\ref{fig:ocs-samples} shows 3 samples in the OCS data set that we used to characterize the performance of the Voigt-1D window. 
From top to bottom, these 3 samples come from the OCS line at 61~GHz, 195~GHz, and 260~GHz, respectively. 
The FID envelope shifts from a Lorentzian-dominated shape at the low frequency end, to almost equally weighted Gaussian-Lorentzian shape at higher frequencies.

The top row shows the raw FID signal, and the time domain fit. 
For all entries, the fit converges well and returns valid $b_0$ parameter. 
The 2\textsuperscript{nd} row shows the magnitude FT of the FID, without and with zero-padding to 20~$\upmu$s.
The 3\textsuperscript{rd} to 4\textsuperscript{th} rows show the result of Voigt-1D windowed spectral lines, without and with zero-padding. 
The 5\textsuperscript{th} to 6\textsuperscript{th} rows show the result of Kaiser windowed spectral lines, without and with zero-padding. 
For all frequency domain spectra, we normalized the peak intensity to unity so that the noise of the residual can be visually compared. 

All spectra without zero-padding are well modeled by Voigt and Gaussian (for Kaiser $\pi\alpha=8$) line profiles.
The fitted residuals are almost flat across the plotted region. 
For spectra with zero-padding, strong baseline ripples are present in un-windowed spectra. 
All window functions are able to erase these ripples. 
The Kaiser ($\pi\alpha=4$ and 8) windows, however, produce lineshapes that slightly deviate from the line profile models, 
as we can find larger residuals around the foot of the lines. 
The Voigt-1D windows still produce perfect Voigt lineshapes, as the residuals are still flat.

\subsection{The \ce{CH3CN} fit\label{section:ch3cn-fit-more}}

\begin{figure}[htp!]
\centering
\includegraphics{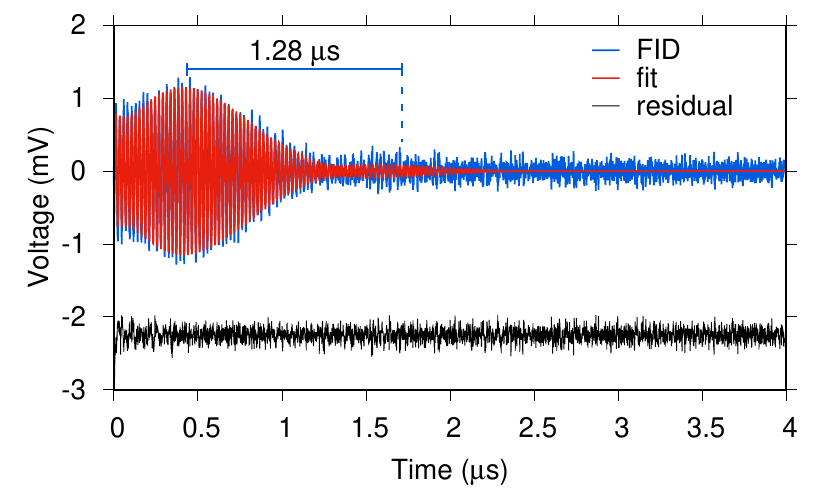}
\caption{The FID of the $J_{10\leftarrow 9}, K=9$ line of \ce{CH3CN} identical to the data shown in Figure~4.
The DAQ delay, however, is set to 450~ns. 
The FID signal is fitted to $f(t) = \exp(-0.4508 t^2 - 1.106(22) t)$. 
The blue trace shows the FID signal,  the red trace shows the least-squared fit, and the black trace shows the fit residual (shifted for clarity). \label{fig:CH3CN-FID-hfs-450}}
\end{figure}

When targeting high SnR, the closely-spaced hyperfine lines of \ce{CH3CN} can cause a pitfall for the Voigt-1D window parameter selection,
because of the interaction of the slowly varying beating envelope of the original FID signal, and the rapid rise of the window function profile. 
To demonstrate the complicated effect, 
we tested the treatment of Voigt-1D window with two FID signals. 
The two FID signals are identical data set, except for the DAQ delay time, one for 256~ns (presented in Figure~4 in the main article), and another for 450~ns (see Figure~\ref{fig:CH3CN-FID-hfs-450}). 
By altering the DAQ delay time, we can evaluate the effect of the node of the beating envelope.
Following the convention in the main article, the Voigt-1D parameter set (1) optimizes for maximum SnR, and set (2) optimizes for maximum SnR/FWHM. 
In addition, we manually fixed the Voigt-1D $b$ parameter to $-b_0$, 
where $b_0$ is the least-square fitted exponential decay coefficient of the FID signal. 
After fixing $b=-b_0$, the $a$ parameter is again optimized for maximum SnR and SnR/FWHM under the condition of the fixed $b$. 
The truncated Kaiser window was also used for comparison.

\begin{figure}[htp!]
\centering
\includegraphics[width=\textwidth]{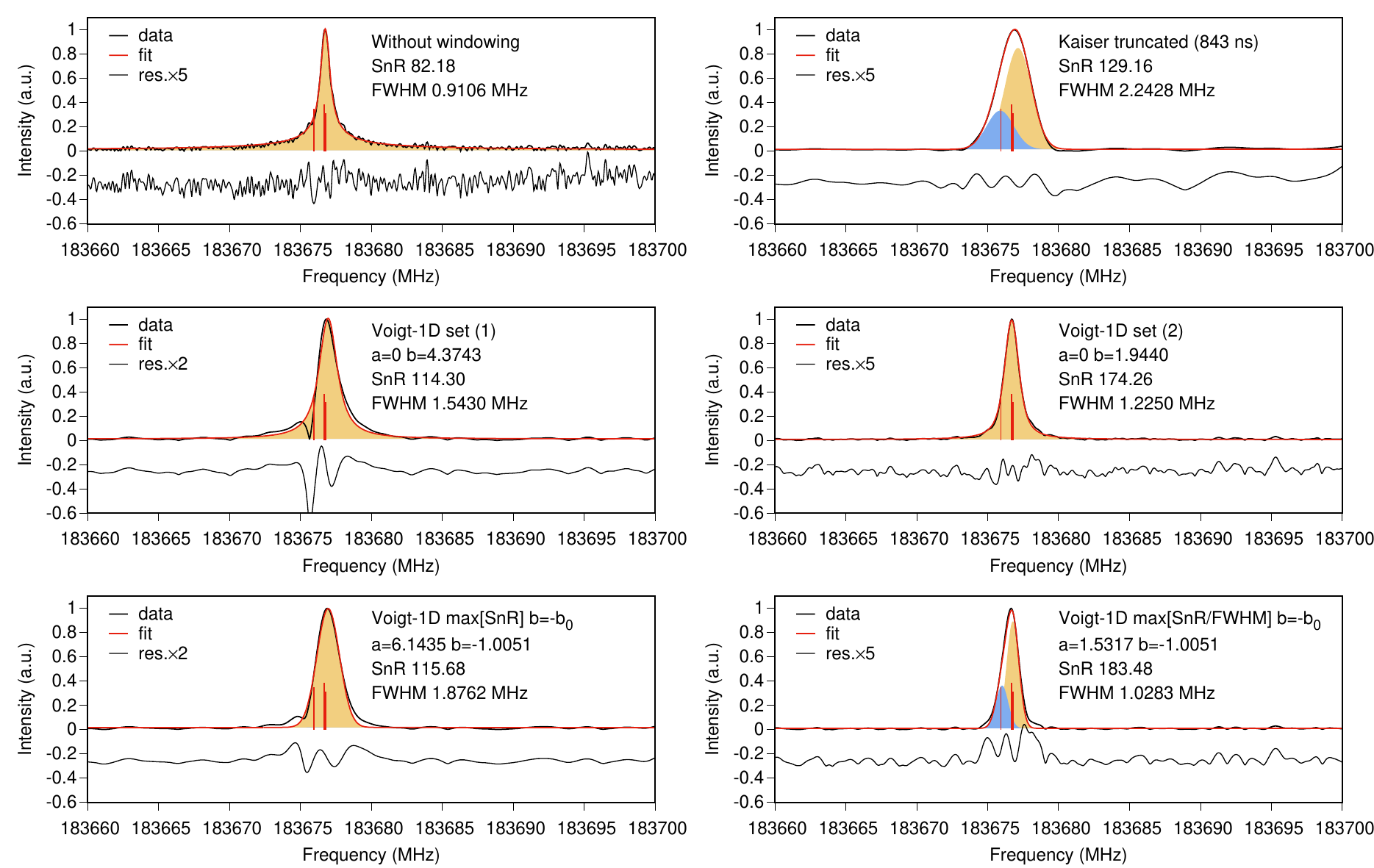}
\caption{Fit results of the $J_{10\leftarrow 9}, K=9$ spectral line of \ce{CH3CN}, with DAQ delay of 256~ns.
The frequency is the intermediate frequency with LO equals to 183720~MHz. 
The spectral intensities are normalized to unity to ease the visual comparison of the noise level. 
The type of window and window parameters are labeled in each subplot. 
The black trace shows the magnitude spectrum, the red trace shows the overall fit, and the fit residual is also plotted below each spectrum.
The solid curves shows the individual components of the fit. The red vertical sticks shows the catalog frequency and intensity of the 3 strong hyperfine components.
The center row shows the Voigt-1D window with parameter set (1) for maximizing SnR, and parameter set (2) for maximizing SnR/FWHM, based on theory.
The bottom row shows the Voigt-1D window with only $a$ optimized for maximizing SnR and SnR/FWHM under the constraint of $b=b_0$. 
$a$ and $b$ are in units of MHz$^2$ and MHz, respectively. 
The noise level is estimated using the fitted residuals outside $\pm3$~MHz of the peaks. 
\label{fig:CH3CN-fit-256}}
\end{figure}

\begin{figure}[htp!]
\centering
\includegraphics[width=\textwidth]{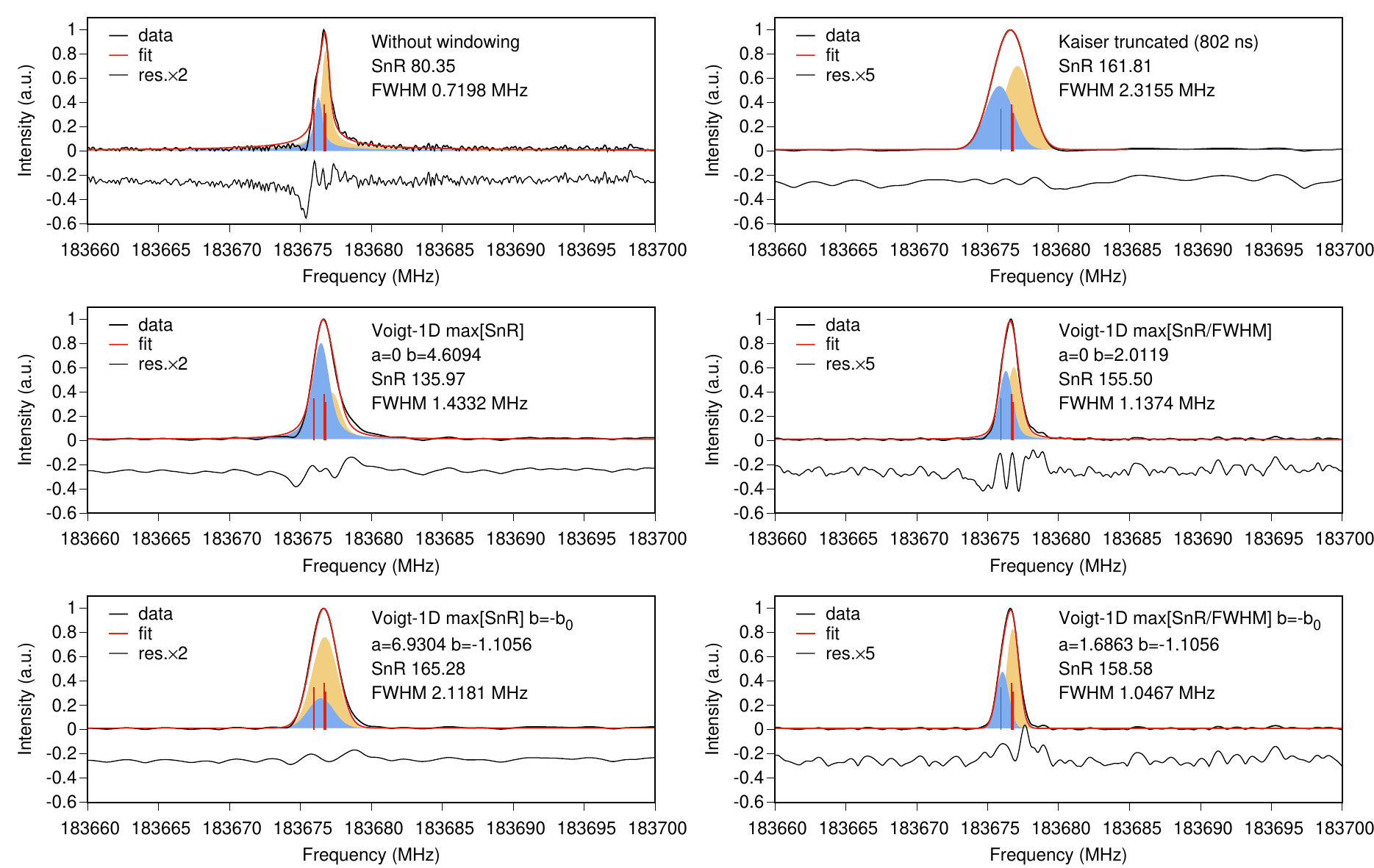}
\caption{Fit results of the $J_{10\leftarrow 9}, K=9$ spectral line of \ce{CH3CN}, with DAQ delay of 450~ns.
The frequency is the intermediate frequency with LO equals to 183720~MHz. 
The spectral intensities are normalized to unity to ease the visual comparison of the noise level. 
The type of window and window parameters are labeled in each subplot. 
The black trace shows the magnitude spectrum, the red trace shows the overall fit, and the fit residual is also plotted below each spectrum.
The solid curves shows the individual components of the fit. The red vertical sticks shows the catalog frequency and intensity of the 3 strong hyperfine components.
The center row shows the Voigt-1D window with parameter set (1) for maximizing SnR, and parameter set (2) for maximizing SnR/FWHM, based on theory.
The bottom row shows the Voigt-1D window with only $a$ optimized for maximizing SnR and SnR/FWHM under the constraint of $b=b_0$. 
$a$ and $b$ are in units of MHz$^2$ and MHz, respectively. 
The noise level is estimated using the fitted residuals outside $\pm3$~MHz of the peaks. 
\label{fig:CH3CN-fit-450}}
\end{figure}

Figure~\ref{fig:CH3CN-fit-256} shows the results of window functions on the FID with 256~ns delay, 
and Figure~\ref{fig:CH3CN-fit-450} shows the results on the FID of 450~ns delay. 
To avoid the influence of imperfection line profile fit to the SnR calculation, 
we only selected the fit residuals outside $\pm3$~MHz of the two peaks for noise estimation. 

The effect of the DAQ delay time is already prominent on the un-windowed spectra. 
For the 256~ns delayed data, only one peak is present. 
For the 450~ns delayed data, the hyperfine structure becomes partially resolvable. 
However, the 450~ns delayed data shows asymmetric lineshape. 

The results of Voigt-1D window set (1) and (2) on the two FID signals all show asymmetric lineshape. 
The parameter set (1), which is optimized for a larger $b$ value, causes more severe lineshape imperfection than the parameter set (2). 
When fitted with the Voigt profile, 
either no hyperfine structure can be resolved (256~ns delayed data), 
or the relative intensities of the two frequency components are far from correctness (450~ns delayed data).
On the contrary, the truncated Kaiser window generates reliable results in both cases, 
with partially resolved frequency components with correct relative intensity. 
The drawback is that the total FWHM of the lines treated by the truncated Kaiser window is quite large. 
Therefore, the hyperfine structure is not really ``resolved'', 
as one peak may also be sufficient to fit the spectra line, 
and therefore the SnR is in fact the sum of two frequency components.

If $b$ is fixed to $-b_0$, the Voigt-1D window then produces spectral lines with almost symmetric lineshape. 
By setting $b=-b_0$, the window function cancels out the exponential decay terms in the FID, most of which is introduced by the DAQ delay. 
Smaller $a$ value also helps the lineshape correction, 
and therefore the result optimized for SnR/FWHM is closer to a perfect Voigt profile. 
The SnR is not as high as what is produced by the truncated Kaiser window, 
but the FWHM is also smaller. 
Therefore, fixing $b=-b_0$ and optimize $a$ for highest SnR/FWHM is a practical choice for obtaining high SnR spectrum 
for two closely spaced lines. 

Although sensitive to the low frequency beating envelope of the FID, 
the examples with $b$ fixed to $-b_0$ demonstrates the wide tunability of the Voigt-1D window function.
There is almost always a suitable range of $a$ and $b$ values in the parameter space
that can produce high SnR spectral lines with correct lineshape. 
The result is not the optimal in theory, but is acceptable for real spectra. 

\bibliographystyle{elsarticle-harv}